\documentclass{azh}

\usepackage{graphicx}
\usepackage{natbib}
\usepackage{lscape}

\def\sun{\hbox{$\odot$}}
\def\degr{\hbox{$^\circ$}}
\def\arcmin{\hbox{$^\prime$}}
\def\arcsec{\hbox{$^{\prime\prime}$}}

\begin{document} 

\title{Studies of the Stellar Populations of Galaxies Using Two-Color Diagrams}

\author{A.~S.~Gusev,$^1$ 
            S.~A.~Guslyakova,$^2$
            A.~P.~Novikova,$^1$ 
            M.~S.~Khramtsova,$^3$
            V.~V.~Bruevich,$^1$
            O.~V.~Ezhkova$^1$}

\institute{$^1$ Sternberg Astronomical Institute, Lomonosov Moscow University, 
Universitetskii pr. 13, Moscow 119992 Russia \\
              $^2$ Space Research Institute, Russian Academy of Sciences,
84/32 Profsoyuznaya Str, Moscow 117997, Russia \\
              $^3$ Institute of Astronomy, Russian Academy of Sciences,
Pyatnitskaya 48, Moscow 119017, Russia}

\date{Received December 12, 2014; in final form, March 13, 2015}
\offprints{Alexander~S.~Gusev, \email{gusev@sai.msu.ru}}

\titlerunning{Stellar Populations of Galaxies}
\authorrunning{Gusev et al.} 

\abstract{Variations in the photometric parameters of stellar systems as a function 
of their evolution and the stellar populations comprising them are investigated. A set 
of seven evolutionary models with an exponential decrease in the star-formation rate 
and 672 models with a secondary burst of star formation are considered. The 
occurrence of a secondary burst of star formation can shift the position of a stellar 
system on two-color diagrams to the right or left of the normal color sequence for 
galaxies and the extinction line. This makes it possible to estimate the composition 
of the stellar population of a galaxy with a nonmonotonic star-formation history from 
its position on two-color diagrams. Surface photometry in both the optical ($UBVRI$) 
and near-IR ($JHK$) is used to study the stellar populations and star-formation 
histories in the structural components (nucleus, bulge, disk, spiral arms, bar, ring) of 
26 galaxies of various morphological types (from S0 to Sd). Components (nucleus, 
bulge, bar) with color characteristics corresponding to stellar systems with secondary 
bursts of star formation are indicated in 10 of the 26 galaxies. The parameters of these 
secondary bursts are estimated. Five of the 10 galaxies with complex star-formation 
histories display clear structural perturbations. Appreciable differences in the 
photometric characteristics of relatively red early-type galaxies (S0--Sb) and relatively 
blue later-type galaxies (Sb--Sd) have been found. Galaxies of both early and late 
types are encountered among the Sb--galaxies. Lenticular galaxies do not display 
different photometric characteristics from early-type spiral galaxies. \\

{\bf DOI:} 10.1134/S1063772915100029 \\

}

\maketitle

\section{INTRODUCTION}

Determining the composition of the stellar populations of galaxies and of their individual 
large-scale elements, such as bulges, disks, nuclei, bars, and spiral arms, is key for 
studies of galactic evolution. Modern receiver instrumentation and methods for reducing 
and storing data make it possible, in principle, to derive the detailed distributions of 
parameters of such stellar populations (their age, metallicity, velocity dispersion) in 
galaxies that cannot be resolved into individual stars. However, this requires joint sets 
of observations, such as surface photometry and spectrophotometry over as wide a 
bandwidth as possible, and one-dimensional and two-dimensional spectroscopy with 
high resolution. Such studies have been successfully carried out for nearby, bright 
galaxies \citep[see, e.g.,][]{gav02,chi09}, but have not been applied to large numbers 
of objects.

Optical surface photometry is the simplest way to obtain information about the structure 
and stellar populations of galaxies. Photometric observations require standard receivers 
and are possible on small (1-2 m) telescopes. Existing databases contain freely available 
images of many hundreds of galaxies in the $UBVRI$ bands, which can also be used 
to analyze photometric results. The extensive database of near-IR ($JHK$) galaxy 
images obtained as part of the 2MASS project \citep{jar00} can be used to supplement 
optical data.

Direct use of color indices -- the main indicators of the composition and age of a stellar 
population -- is hindered by uncertainties associated with the chemical composition and 
amount of interstellar absorption in a given galaxy. Unfortunately, the shift of a stellar 
system in two-color diagrams due to the effects of metallicity and reddening are along 
the normal color sequence (NCS) of galaxies \citep{but95,sch96,zas83} -- a sequence 
of stellar systems with standard star-formation histories; i.e., with star-formation rates 
(SFRs) that fall off exponentially with time \citep{san86}. In general, determining the 
metallicity and absorption in a galaxy requires spectroscopic observations, which, in 
turn, require more expensive and difficult to use equipment and time on larger 
telescopes.

\begin{table*}
\begin{center}
\caption{Main characteristics of the galaxies.}
\label{table:tab1}
\begin{tabular}{|l|c|c|c|c|c|c|c|c|c|c|c|}
\hline\hline 
Galaxy & \multicolumn{2}{c|}{Type} & $m_B$ & $M_B^{0, i}$ & $i$ & PA & $D_{25}$ & 
$D_{25}$, kpc & $d$, Mpc & $A(B)_{G}$ & $A(B)_{i}$ \\
\hline
NGC 245    & SA(rs)b        & 3.1   & 12.72$^m$ & --21.12$^m$ & $21\degr$ & $145\degr$ & 
$1.23\arcmin$ & 19.3 & 53.8 & 0.097$^m$ & 0.04$^m$ \\
NGC 266    & SB(rs)ab       & 1.6   & 12.27  & --21.94 & 15 & 95  & 3.09  & 57.4 & 63.8 & 0.252 & 0.01 \\
NGC 524    & SA0+(rs)      & --1.2 & 11.01  & --21.72 & 5   & 55  & 3.63  & 34.1 & 32.3 & 0.299 & 0.00 \\
NGC 532    & Sab?             & 1.9   & 13.10  & --19.59 & 71 & 31  & 2.29  & 21.1 & 31.7 & 0.337 & 0.42 \\
NGC 628    & SA(s)c          & 5.2   & 9.70    & --20.72 & 7   & 25  & 10.46 & 21.9 & 7.2  & 0.254 & 0.04 \\
NGC 783    & Sc                & 5.3    & 13.18  & --22.01 & 43 & 57  & 1.42  & 29.1 & 70.5 & 0.222 & 0.45 \\
NGC 1138   & SB0             & --1.9 & 13.31  & --19.49 & 35 & 95  & 2.04  & 19.6 & 33.0 & 0.542 & 0.00 \\
NGC 1589   & Sab              & 2.4   & 11.80  & --21.86 & 80 & 159 & 3.16  & 45.5 & 49.5 & 0.307 & 0.61 \\
NGC 2336   & SAB(r)bc     & 4.0   & 11.19  & --22.14 & 55 & 175 & 5.02  & 47.0 & 32.2 & 0.120 & 0.41 \\
NGC 3184   & SAB(rs)cd    & 5.9   & 10.31  & --19.98 & 14 & 117 & 7.59  & 22.5 & 10.2 & 0.060 & 0.02 \\
NGC 3726   & SAB(r)c       & 5.1    & 10.31  & --20.72 & 49 & 16  & 5.25  & 21.9 & 14.3 & 0.060 & 0.30 \\
NGC 4136   & SAB(r)c       & 5.2    & 11.92  & --18.38 & 22 & 30  & 2.40  & 5.6  & 8.0   & 0.066 & 0.05 \\
NGC 5351   & SA(r)b?        & 3.1   & 12.57  & --21.16 & 60 & 101 & 2.40  & 35.6 & 51.1 & 0.074 & 0.40 \\
NGC 5585   & SAB(s)d       & 6.9   & 10.94  & --18.73 & 53 & 34  & 4.27  & 7.1  & 5.7   & 0.057 & 0.38 \\
NGC 5605   & (R)SAB(rs)c & 4.9   & 12.58  & --20.86 & 36 & 65  & 1.62  & 21.1 & 44.8 & 0.318 & 0.15 \\
NGC 5665   & SAB(rs)c      & 5.0   & 12.25  & --20.42 & 53 & 151 & 1.91  & 17.3 & 31.1 & 0.091 & 0.35 \\
NGC 6217   & (R)SB(rs)bc  & 4.0   & 11.89  & --20.45 & 33 & 162 & 2.30  & 13.8 & 20.6 & 0.158 & 0.22 \\
NGC 6340   & SA0/a(s)       & 0.4   & 11.73  & --20.03 & 20 & 155 & 3.16  & 18.6 & 20.2 & 0.173 & 0.02 \\
NGC 6946   & SAB(rs)cd    & 5.9   & 9.75    & --20.68 & 31 & 62  & 15.48 & 26.6 & 5.9  & 1.241 & 0.04 \\
NGC 7280   & SAB0+(r)      & --1.3 & 12.82  & --19.44 & 59 & 76  & 2.09  & 15.8 & 25.9 & 0.201 & 0.00 \\
NGC 7331   & SA(s)b          & 3.9   & 10.20  & --21.68 & 75 & 169 & 9.78  & 40.1 & 14.1 & 0.331 & 0.61 \\
NGC 7351   & SAB0?          & --2.1 & 13.39  & --17.22 & 76 & 2    & 1.66  & 5.7  & 11.7 & 0.161 & 0.00 \\
NGC 7678   & SAB(rs)c      & 4.9    & 12.50  & --21.55 & 44 & 21  & 2.08  & 28.9 & 47.8 & 0.178 & 0.23 \\
NGC 7721   & SA(s)c          & 4.9    & 11.11  & --21.18 & 81 & 16  & 3.02  & 23.1 & 26.3 & 0.121 & 0.98 \\
IC 1525       & SBb              & 3.1    & 12.51  & --21.89 & 48 & 27  & 1.95  & 39.5 & 69.6 & 0.410 & 0.24 \\
UGC 11973 & SAB(s)bc      & 3.9   & 13.34   & --22.47 & 81 & 39 & 3.46  & 59.2 & 58.8 & 0.748 & 0.85 \\[1mm]
\hline
\end{tabular}
\end{center}
\end{table*} 

The goal of our current study is to elucidate what properties of the stellar populations 
of galaxies can be determined using optical ($UBVRI$) and infrared ($JHK$) 
photometry in the absence of independent information about absorption and 
chemical composition. As observational material, we used previously obtained 
surface photometry data for galaxies of various morphological types, both published 
and unpublished (Section~2). Section~3 considers the variations of the positions of 
stellar systems on two-color diagrams as a function of their particular star formation 
histories, using modern evolutionary models. Section~4 compares the photometric 
data and models in order to estimate the stellar populations in various components of 
the galaxies. We discuss the results in Section~5, and formulate our conclusions in 
Section~6.

\section{GALAXY SAMPLE}

We chose 26 galaxies with various morphological types for which $UBVRI$ and $JHK$ 
photometry was available for this study. Table~\ref{table:tab1} presents the main 
information about the galaxies -- morphological type (in letter and numerical form); 
apparent magnitude $m_B$ corrected for Galactic absorption and absorption due to the 
inclination of the galaxy; absolute magnitude $M_B^{0, i}$; inclination $i$; position 
angle PA and diameter at the $25^m$ isophote in the $B$ band $D_{25}$, taking into 
account Galactic absorption and absorption due to the inclination of the galaxy; distance 
$d$; Galactic absorption $A(B)_{G}$; and absorption due to the inclination of the 
galaxy $A(B)_{i}$. The data on the morphological types (in letter form) and the 
absorption arising in our Galaxy, $A(B)_{G}$, were taken from the NASA Extragalactic 
Database (NED), and the remaining parameters from the HyperLeda database. There was 
no information about the position angles of some galaxies in the HyperLeda database; 
in these cases, the PAs were taken from NED.

The observations, reduction, and results of the surface-photometry data for 18 of the galaxies 
are described in 
\citet{art97,art99,art00,gus02a,gus02b,gus03a,gus03b,gus04,gus06a,gus06b,bru07,bru10,bru11} 
(Table~\ref{table:tab2}). Here, we use these earlier results, taking into account the new data of 
\citet{sch11} on the distribution of absorption in our Galaxy.

The data for eight galaxies (NGC~245, NGC~266, NGC~6340, NGC~6946, NGC~7331, 
NGC~7351, NGC~7721, and UGC~11973) have not been published earlier. We describe 
the observations and reduction for these galaxies, as well as for NGC~6217 and NGC~7678, 
which we have observed multiple times, below in this section.

The observations of the galaxies were carried out in 2002--2006 on the 1.5~m telescope of 
the Maidanak Observatory of the Institute of Astronomy of the Academy of Sciences of the 
Republic of Uzbekistan (focus length 12~m), using a SITe-2000 CCD array. With the 
broadband $U$, $B$, $V$, $R$, and $I$ filters used, the CCD array realizes a photometric 
system close to the standard Johnson--Cousins $UBVRI$ system. A detailed description of 
the characteristics of the telescope and the instrumental photometric system can be found in 
\citet{art10}. The CCD array was cooled by liquid nitrogen, and was $2000\times800$ pixels 
in size, providing a field of view of $8.9\arcmin\times3.6\arcmin$ and an image scale of 
$0.267\arcsec$/pixel. Table~\ref{table:tab3} presents a journal of these observations.

\begin{table*}
\begin{center}
\caption{General information about the observations.}
\label{table:tab2}
\begin{tabular}{|l|c|c|c|c|c|}
\hline\hline 
Galaxy & Filter & Year & Telescope & Observatory & Reference \\
\hline
NGC 245    & $UBVRI$        & 2005          & 1.5 m               & Maidanak          & \\
NGC 266    & $UBVRI$        & 2005          & 1.5 m               & Maidanak          & \\
NGC 524    & $UBVRIJHK$ & 2003          & 1.5 m                & Maidanak          & \citet{gus06a,gus06b} \\
NGC 532    & $UBVRIJHK$ & 2003          & 1.5 m                & Maidanak          & \citet{gus06a,gus06b} \\
NGC 628    & $UBVRI$        & 2002          & 1.5 m                & Maidanak          & \citet{bru07} \\
NGC 783    & $UBVRIJHK$  & 2003          & 1.5 m               & Maidanak          & \citet{gus06a,gus06b} \\
NGC 1138   & $UBVRIJHK$ & 2003          & 1.5 m               & Maidanak          & \citet{gus06a,gus06b} \\
NGC 1589   & $UBVRIJHK$ & 2003          & 1.5 m               & Maidanak          & \citet{gus06a,gus06b} \\
NGC 2336   & $UBVRIJHK$ & 2001          & 1.8 m               & Bohyunsan        & \citet{gus03b} \\
NGC 3184   & $BVRI$          & 1998, 1999 & 1 m, 1 m (JKT) & SAO, La Palma & \citet{gus02b} \\
NGC 3726   & $BVRI$          & 1998          & 1 m                  & SAO                 & \citet{gus02a} \\
NGC 4136   & $BVRIJHK$   & 1998           & 1 m                  & SAO                 & \citet{gus03a} \\
NGC 5351   & $BVRIJHK$   & 1998           & 1 m                  & SAO                 & \citet{gus04} \\
NGC 5585   & $UBVRI$       & 2005           & 1.5 m               & Maidanak          & \citet{bru10} \\
NGC 5605   & $BVRI$          & 1997          & 1.5 m                & Maidanak          & \citet{art00} \\
NGC 5665   & $BVRI$          & 1997          & 1.5 m                & Maidanak          & \citet{art00} \\
NGC 6217   & $UBVRI$       & 1988, 1989, & 1 m,                 & Maidanak          & \citet{art99} \\
                   &                       & 1997, 2006  & 1.5 m               &                          & \\
NGC 6340   & $UBVRI$       & 2006           & 1.5 m               & Maidanak          & \\
NGC 6946   & $UBVRI$       & 2002           & 1.5 m               & Maidanak          & \\
NGC 7280   & $UBVRIJHK$ & 2003          & 1.5 m                & Maidanak          & \citet{gus06a,gus06b} \\
NGC 7331   & $UBVRI$       & 2005           & 1.5 m               & Maidanak          & \\
NGC 7351   & $UBVRI$       & 2005           & 1.5 m               & Maidanak          & \\
NGC 7678   & $UBVRI$       & 1989, 2005  & 1 m, 1.5m        & Maidanak          & \citet{art97} \\
NGC 7721   & $UBVRI$       & 2005           & 1.5 m               & Maidanak          & \\
IC 1525       & $UBVRI$       & 2005           & 1.5 m               & Maidanak          & \citet{bru11} \\
UGC 11973 & $UBVRI$       & 2005           & 1.5 m               & Maidanak          & \\[1mm]
\hline
\end{tabular}
\end{center}
\end{table*}
 
The subsequent reduction and calibration of the data were carried out using standard 
procedures in the ESO--MIDAS image-reduction system, described in detail, for example, 
in \citet{bru07,bru10}.

To investigate the IR properties of the galaxies, we used data from the 2MASS catalog, which 
contains $J$, $H$, and $K$ images of the galaxies. These images were reduced using a similar 
procedure. The seeing for these $JHK$ images is $1\arcsec$, with an image scale of 
$1.0\arcsec$/pixel $\times 1.0\arcsec$/pixel. In contrast to the data of \citet{gus06a,gus06b}, 
where we used the 2MASS $JHK$ photometric system, here, we converted the data to the 
$JHK$ system of \citet{bes88}, in accordance with citet{car01,kin02}: 
$K_{\rm BB} = K_{\rm 2MASS}+0.044$, 
$(J-H)_{\rm BB} = 1.020(J-H)_{\rm 2MASS}+0.046$, 
$(J-K)_{\rm BB} = 1.029(J-K)_{\rm 2MASS}+0.011$, and 
$(H-K)_{\rm BB} = 1.004(H-K)_{\rm 2MASS}-0.028$.

We constructed color equations and corrected for atmospheric extinction using $U$, $B$, $V$, 
$R$, and $I$ observations of the Landolt standard stars PG~0231+051, PG~2213--006, 
PG~2331+055, SA~92, SA~110, SA~111, SA~113, and SA~114 \citep{lan92} obtained on the 
same night over a broad range of air masses. When possible, we compared the data obtained 
with the results of aperture photometry data for the galaxies contained in the electronic version 
of the HyperLeda catalog. According to our estimates, the photometric calibration errors do not 
exceed $0.02^m-0.07^m$ in $B$, $V$, $R$, and $I$, and $0.1^m$ in $U$.

All the data were corrected for Galactic absorption and absorption arising due to the inclination 
of the galaxy. We took the Hubble constant to be H$_0 = 75$ km$\cdot$s$^{-1}$Mpc$^{-1}$. 
The linear scales of the images are presented in Table~\ref{table:tab3}.

\section{MODELS}

\subsection{Models with an Exponential Decrease in the Star-Formation Rate}

The composition of the stellar population of a galaxy at any given time can be described using 
two parameters: the initial mass function (IMF) and the star-formation history, i.e., the variation 
in the SFR with time.

\begin{table*}
\begin{center}
\caption{Journal of observations.}
\label{table:tab3}
\begin{tabular}{|l|c|c|c|c|c|c|}
\hline\hline 
Galaxy & Date & Filter & Exposure, & Air mass & Seeing & Scale,   \\
       &      &        & s         &          &        & pc/pixel \\
\hline
NGC 245 & 1/2.11.2005 & $U$ & 2$\times$300 & 1.31 & $1.1\arcsec$ & 69.6 \\
& & $B$ & 2$\times$240 & 1.31 & 1.0 & \\
& & $V$ & 2$\times$180 & 1.31 & 1.1 & \\
& & $R$ & 2$\times$90 & 1.31 & 0.8 & \\
& & $I$ & 2$\times$90 & 1.32 & 0.8 & \\
\hline
NGC 266 & 3/4.11.2005 & $U$ & 2$\times$300 & 1.01 & 1.0 & 82.5 \\
& & $B$ & 2$\times$240 & 1.01 & 0.9 & \\
& & $V$ & 2$\times$180 & 1.01 & 0.8 & \\
& & $R$ & 2$\times$120 & 1.01 & 0.7 & \\
& & $I$ & 2$\times$90 & 1.01 & 0.7 & \\
\hline
NGC 6217 & 21/22.06.2006 & $U$ & 2$\times$300 & 1.39 & 1.7 & 26.6 \\
& & $B$ & 2$\times$180 & 1.40 & 1.4 & \\
& & $V$ & 2$\times$120 & 1.40 & 1.4 & \\
& & $R$ & 3$\times$90 & 1.41 & 1.2 & \\
& & $I$ & 1$\times$60 & 1.42 & 1.1 & \\
\hline
NGC 6340 & 19/20.06.2006 & $U$ & 2$\times$300 & 1.29 & 1.2 & 26.1 \\
& & $B$ & 2$\times$240 & 1.30 & 0.9 & \\
& & $V$ & 2$\times$120 & 1.31 & 0.8 & \\
& & $R$ & 2$\times$60 & 1.28 & 0.8 & \\
& & $I$ & 2$\times$60 & 1.28 & 0.7 & \\
\hline
NGC 6946 & 12/13.09.2002 & $U$ & 4$\times$300 & 1.28 & 1.6 & 7.6 \\
(northern & & $B$ & 3$\times$300 & 1.33 & 1.3 & \\
part) & & $V$ & 2$\times$240 & 1.38 & 1.3 & \\
& & $R$ & 2$\times$180 & 1.23 & 1.0 & \\
& & $I$ & 2$\times$150 & 1.24 & 0.9 & \\
& 21/22.10.2002 & $U$ & 4$\times$300 & 1.18 & 1.1 & \\
& & $B$ & 1$\times$300 & 1.20 & 1.0 & \\
& & $V$ & 2$\times$240 & 1.22 & 0.9 & \\
& & $R$ & 2$\times$180 & 1.15 & 0.8 & \\
& & $I$ & 2$\times$150 & 1.16 & 0.8 & \\
\hline
NGC 6946 & 12/13.09.2002 & $U$ & 4$\times$300 & 1.14 & 1.4 & 7.6 \\
(southern & & $B$ & 2$\times$300 & 1.17 & 1.1 & \\
part) & & $V$ & 2$\times$240 & 1.19 & 1.1 & \\
& & $R$ & 2$\times$180 & 1.12 & 0.7 & \\
& & $I$ & 2$\times$150 & 1.13 & 0.8 & \\
& 21/22.10.2002 & $U$ & 4$\times$300 & 1.10 & 1.1 & \\
& & $B$ & 2$\times$300 & 1.12 & 1.3 & \\
& & $V$ & 2$\times$240 & 1.14 & 0.9 & \\
& & $R$ & 2$\times$180 & 1.09 & 1.2 & \\
& & $I$ & 2$\times$150 & 1.09 & 0.9 & \\
\hline
NGC 7331 & 8/9.09.2005 & $U$ & 8$\times$300 & 1.03 & 1.4 & 18.2 \\
(central & & $B$ & 2$\times$240 & 1.02 & 1.3 & \\
part) & & $V$ & 2$\times$150 & 1.03 & 1.3 & \\
& & $R$ & 2$\times$90 & 1.00 & 1.0 & \\
& & $I$ & 2$\times$50 & 1.01 & 1.0 & \\
& 10/11.09.2005 & $U$ & 4$\times$300 & 1.01 & 1.0 & \\
& & $B$ & 2$\times$300 + 1$\times$150 & 1.00 & 1.0 & \\
& & $V$ & 4$\times$150 & 1.00 & 1.0 & \\
& & $R$ & 2$\times$90 & 1.01 & 0.9 & \\
& & $I$ & 2$\times$60 & 1.01 & 0.8 & \\
\hline
NGC 7351 & 2/3.11.2005 & $U$ & 2$\times$300 & 1.37 & 1.5 & 15.1 \\
& & $B$ & 2$\times$180 & 1.37 & 1.3 & \\
& & $V$ & 2$\times$180 & 1.38 & 1.3 & \\
& & $R$ & 2$\times$90 & 1.39 & 1.0 & \\
& & $I$ & 2$\times$90 & 1.39 & 1.0 & \\[1mm]
\hline
\end{tabular}
\end{center}
\end{table*}

\setcounter{table}{2}
\begin{table*}
\begin{center}
\caption{Continued}
\begin{tabular}{|l|c|c|c|c|c|c|}
\hline\hline
Galaxy & Date & Filter & Exposure, & Air mass & Seeing & Scale,   \\
       &      &        & s         &          &        & pc/pixel \\
\hline
NGC 7678 & 13/14.09.2005 & $U$ & 5$\times$300 & 1.08 & 1.1 & 61.8 \\
& & $B$ & 2$\times$300 & 1.20 & 1.1 & \\
& & $V$ & 1$\times$240 & 1.05 & 0.9 & \\
& & $R$ & 2$\times$120 & 1.05 & 0.8 & \\
& & $I$ & 2$\times$90 & 1.22 & 1.1 & \\
\hline
NGC 7721 & 3/4.11.2005 & $U$ & 2$\times$300 & 1.42 & 1.2 & 34.0 \\
& & $B$ & 2$\times$240 & 1.42 & 1.3 & \\
& & $V$ & 2$\times$180 & 1.42 & 1.0 & \\
& & $R$ & 2$\times$120 & 1.42 & 0.9 & \\
& & $I$ & 2$\times$90 & 1.42 & 0.9 & \\
\hline
UGC 11973 & 31.10/1.11.2005 & $U$ & 2$\times$300 & 1.00 & 1.7 & 76.0 \\
& & $B$ & 2$\times$240 & 1.01 & 1.3 & \\
& & $V$ & 2$\times$180 & 1.01 & 1.2 & \\
& & $R$ & 2$\times$90 & 1.01 & 1.0 & \\
& & $I$ & 2$\times$90 & 1.01 & 1.0 & \\[1mm]
\hline
\end{tabular}
\end{center}
\end{table*}

The parameters of the IMF only weakly influence the photometric characteritics of the stellar 
population: the most massive stars make a substantial contribution to the radiation, but their 
lifetimes ($\sim1-10$~Myr) are short compared to the characteristic time scale for galactic 
evolution ($\sim1-10$~Gyr), and their contribution to the total mass is too small to 
appreciably increase themetallicity of the interstellar medium. Low-mass stars make an extremely 
small contribution to the radiation of the stellar system in the optical. Small variations in the IMF 
likewise do not give rise to substantial variations in the color characteristics \citep{bru11}. In 
our current study, we used a classical \citet{sal55} IMF with lower and upper mass limits of 
$0.1M\sun$ and $100M\sun$.

Estimates of the variation of the SFR with time are very uncertain. Thirty years ago, \citet{san86} 
proposed a model in which the SFR decreased exponentially with time, with the characteristic 
time scale $\tau$ increasing along the Hubble sequence of galaxies, from ellipticals to late-type 
spirals. The physical meaning behind this dependence was constancy of the star-forming 
efficiency (SFE=SFR/$M_{\rm gas}$, where $M_{\rm gas}$ is the mass of gas in the galaxy) 
with time in each isolated galaxy.

\begin{figure}
\vspace{6mm}
\centerline{\includegraphics[width=8.5cm]{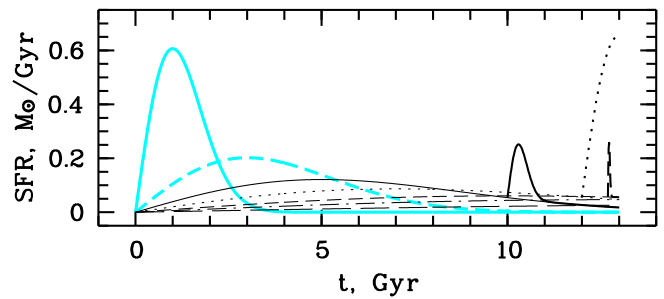}}
\caption{Variation in the SFR as a function of galaxy age. The thick cyan and thin black curves show 
the dependence SFR($t$) according to Eq.~(1) with $\tau_1 = 1, 3, 5, 7, 10, 13$, and 20 Gyr. The 
maxima of SFR($t$) for various values of $\tau_1$ are reached for galaxy ages $t = \tau_1$. The 
thick black curves show secondary bursts of star formation in the stellar systems with 
$\tau_1 = 5$~Gyr 3~Gyr ago with $M_2/M_1 = 0.1$ and $\tau_2 = 300$~Myr (solid); with 
$\tau_1 = 7$~Gyr 1~Gyr ago with $M_2/M_1 = 1$ and $\tau_2 = 1$~Gyr (dotted); and with 
$\tau_1 = 10$~Gyr 300~Myr ago with $M_2/M_1 = 0.01$ and $\tau_2 = 30$~Myr (dashed), in 
accordance with Eq.~(2).
\label{figure:fig1}}
\end{figure}

Later, \citet{gav02} modified the model of Sandage, and suggested instead the dependence
\begin{eqnarray}
{\rm SFR}_1(t, \tau_1) = t/\tau_1^2 \exp(-t^2/2\tau_1^2),
\end{eqnarray}
where $t$ is the time since the birth of the galaxy in Gyr, $\tau_1$ is the characteristic time for 
the decrease in the SFR in Gyr, and the SFR is expressed in $M\sun/10^9$ yr 
(Fig.~\ref{figure:fig1}).

We used the PEGASE.2 program \citep{fio97} to construct evolutionary models for stellar 
systems. Most importantly, we calculated the color indices for stellar systems according to 
Eq.~(1), with characteristic time scales for the decrease in the SFR $\tau_1$ = 1, 3, 5, 7, 10, 
13, and 20 Gyr (Figs.~\ref{figure:fig1}, \ref{figure:fig2}). To reduce the number of free 
parameters, we took the age of the galaxies to be $T = 13$~Gyr and the initial metallicity to 
be $Z(t=0)=0$. Figs.~\ref{figure:fig2}a-\ref{figure:fig2}c show that, overall, the model stellar 
systems with the chosen $\tau_1$, $T$, and $Z(t = 0)$ values correspond to the NCS for 
galaxies derived from the observations of \citet{but95,sch96}. The metallicities of the 
interstellar medium for an age of 13 Gyr calculated for such systems also agrees well with 
observations: the gas metallicities of systems with $\tau_1\le13$~Gyr are $0.5-2 Z\sun$, 
decreasing to $0.2 Z\sun$ only for systems with $\tau_1=20$~Gyr.

\subsection{Models with Secondary Bursts of Star Formation}

\begin{figure*}
\vspace{8mm}
\centerline{\includegraphics[width=16cm]{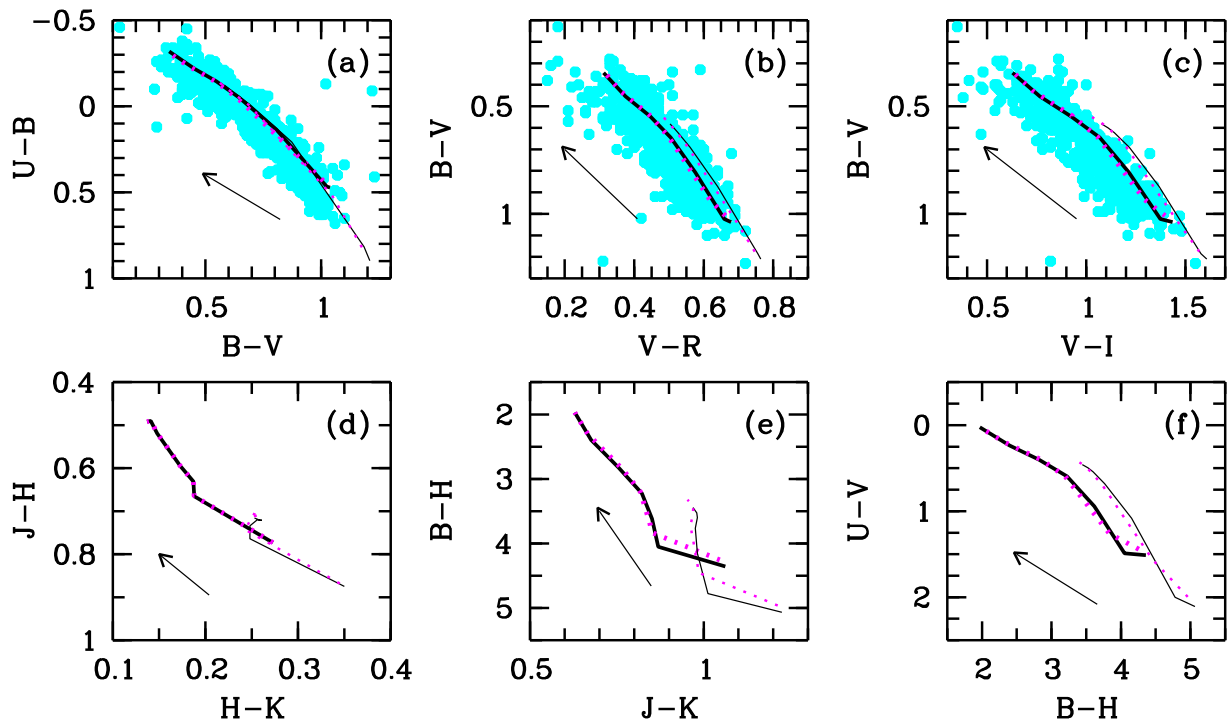}}
\caption{Normal color sequences for galaxies with ages of 13 (solid curves) and 10 Gyr 
(dotted curves) with initial metallicities $Z(t = 0) = 0$ (thick curves) and $Z(t = 0) = Z\sun = 0.02$ 
(thin curves), according to Eq.~(1), on the two-color diagrams {\bf (a)} $(U-B)\div(B-V)$, 
{\bf (b)} $(B-V)\div(V-R)$, {\bf (c)} $(B-V)\div(V-I)$, {\bf (d)} $(J-H)\div(H-K)$, {\bf (e)} 
$(B-H)\div(J-K)$, and {\bf (f)} $(U-V)\div(B-H)$. Stellar systems with larger $\tau_1$ values are 
located further to the upper left in the diagrams. The cyan circles (many merging together) on the 
optical diagrams show the colors of galaxies from the Virgo cluster according to the aperture 
photometry data of \citet{sch96}. The arrows show the variations of the color indices upon 
correction for absorption equal to $A(V) = 1.0^m$.
\label{figure:fig2}}
\end{figure*}

Obviously, we cannot estimate the composition of the stellar populations of stellar systems with 
a standard star-formation history lying along the NCS without some additional data on 
absorption (Fig.~\ref{figure:fig2}). We are interested in systems whose positions in the two-color 
diagrams deviate from the NCS (or from the absorption line). We therefore considered stellar 
systems with secondary bursts of star formation. Such secondary bursts are possible in the 
evolution of galaxies as a whole (e.g., as a result of mergers or the absorption of smaller galaxies), 
as well as their individual components. Components such as disks and active nuclei, where the 
condition ${\rm SFE} = const$ may not be valid, are of special interest.

We used the PEGASE.2 program to construct models for stellar systems with secondary bursts of 
star formation occurring $T_2$ Gyr ago, using the formula
\begin{eqnarray}
{\rm SFR} = {\rm SFR}_1(t, \tau_1) \nonumber
\end{eqnarray} 
for $t < 13 - T_2$ Gyr and
\begin{eqnarray}
{\rm SFR}(t, \tau_1, \tau_2, T_2, M_2/M_1) = {\rm SFR}_1+ \nonumber \\
+(M_2/M_1)(t-13+T_2)/\tau_2^2 \exp[-(t-13+T_2)^2/2\tau_2^2]
\end{eqnarray} 
for $t \ge 13 - T_2$ Gyr, where $\tau_2$ is the characteristic time scale for the decay of the 
secondary burst of star formation in billions of years and $M_2/M_1$ is the mass of gas 
participating in the secondary burst of star formation divided by the total mass of the system at time 
$t = 0$. Fig.~\ref{figure:fig2} presents plots of SFR($t$) for such systems as examples.

We constructed stellar systems with secondary bursts of star formation having the following 
parameters: \\
$\tau_2 = \{1, 10, 100, 1000\}$~Myr, \\
$T_2 = \{10, 30, 100, 300, 1000, 3000\}$~Myr, \\
$M_2/M_1 = \{0.01, 0.1, 1, 10\}$. \\
The metallicity of the system at the onset of the
secondary burst of star formation corresponds to the metallicity of the interstellar medium of the 
evolved system at a time $t - 13 + T_2$~Gyr. We considered the maximum ratio 
$M_2/M_1 = 10$ based on the fact that the fraction of newly formed stars in localized regions 
of a galaxy, such as star-forming complexes, can appreciably exceed the fraction of relative old 
stars in the disk.

The color indices of a stellar system depend on its geometrical shape: the absorption due to dust 
will be different for spheroidal and disk galaxies. For stellar systems with $\tau_1 \ge 3$~Gyr and 
for all secondary bursts of star formation, we used models with disks viewed face-on. For 
systems with $\tau_1 = 1$~Gyr -- the star-formation decay time characteristic of elliptical 
galaxies -- we used spherical models.

\begin{figure*}
\vspace{8mm}
\centerline{\includegraphics[width=16cm]{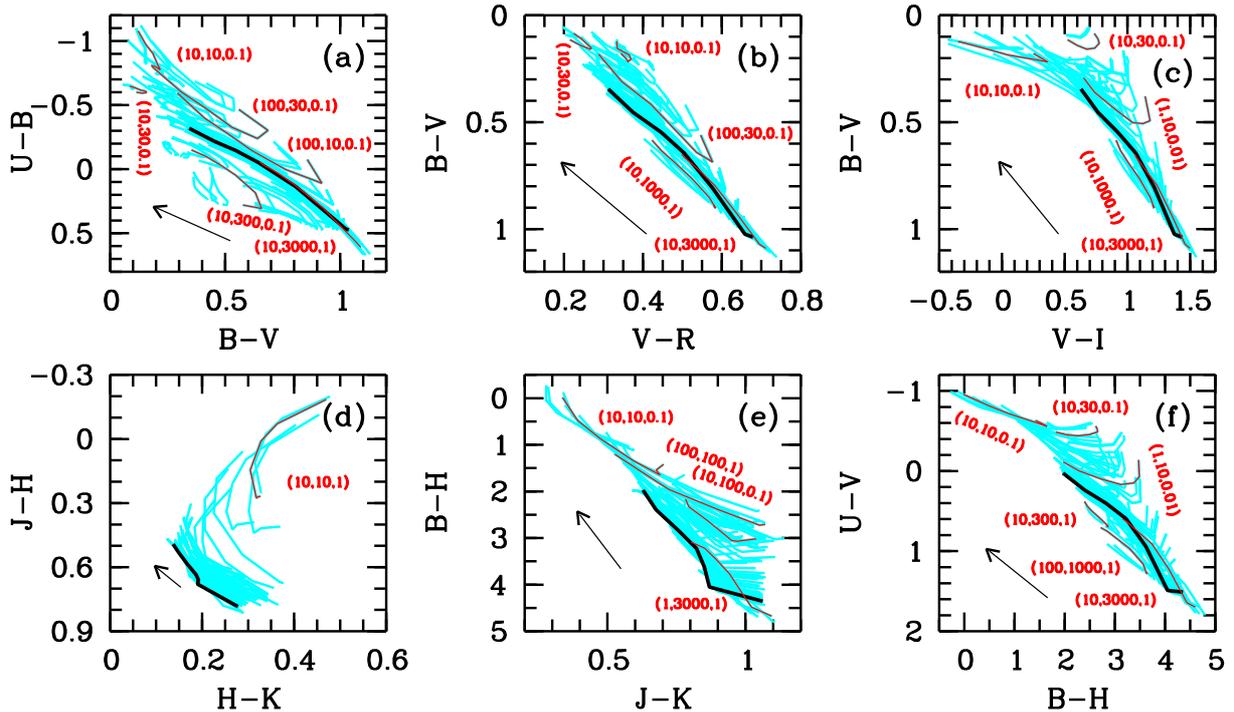}}
\caption{Variations in the two-color diagrams of the NCS for a galaxy with age 13 Gyr and zero 
initial metallicity (thick dark curve) in the case of a secondary burst of star formation (cyan thin 
curves). The red curves show examples of the NCSs of stellar systems with secondary bursts of 
star formation (the numbers in parantheses denote the values of $\tau_2$ in Myr, $T_2$ in Myr, 
and $M_2/M_1$). The remaining notation is as in Fig.~\ref{figure:fig2}.
\label{figure:fig3}}
\end{figure*}

Thus, we calculated the photometric parameters of seven stellar systems with standard star 
formation and 672 ($7\times4\times6\times4$) systems with secondary bursts of star formation.

Let us first consider the positions of stellar systems with a standard star-formation history on the 
two-color diagrams. Fig.~\ref{figure:fig2} presents the NCSs of such systems on $(U-B)\div(B-V)$, 
$(B-V)\div(V-R)$, $(B-V)\div(V-I)$, $(J-H)\div(H-K)$, $(B-H)\div(J-K)$, and $(U-V)\div(B-H)$ 
two-color diagrams. These show that possible variations in the ages of the galaxies do not influence 
their positions in these diagrams: the NCSs for stellar systems with ages of 10 and 13 billion years 
are superposed on one another. The initial metallicity of the interstellar medium has a much greater 
influence: stellar systems with the solar initial metallicity are located toward the lower right in the 
diagrams (i.e., they have redder colors) than systems with zero initial metallicity. Unfortunately, 
in most of the diagrams, variations in the metallicity shift the stellar systems along the absorption 
line. Small shifts from the absorption line due to variations in the chemical composition are 
observed only on the $(U-B)\div(B-V)$ and $(B-H)\div(J-K)$ diagrams (Figs.~\ref{figure:fig2}a, 
\ref{figure:fig2}e). The $J-K$ color index is known as a main indicator of the metallicity of a galaxy 
\citep{bot90}.

The color indices of stellar systems with zero initial metallicity are in fairly good agreement with the 
observed colors of galaxies (Figs.~\ref{figure:fig2}a-\ref{figure:fig2}c). The sequence of stellar 
systems with non-zero initial metallicity deviates from the observed NCS of galaxies. This is 
especially clearly visible on the $(B-V)\div(V-I)$ diagram (Fig.~\ref{figure:fig2}c).

The reddest galaxies (with $U-B > +0.5$) do not agree with the models for stellar systems with a 
standard star formation and zero initial metallicity (Fig.~\ref{figure:fig2}a). We show below that the 
observations of such galaxies are consistent with models with zero initial metallicity and a secondary 
burst of star formation.

\begin{figure*}
\vspace{8mm}
\centerline{\includegraphics[width=16cm]{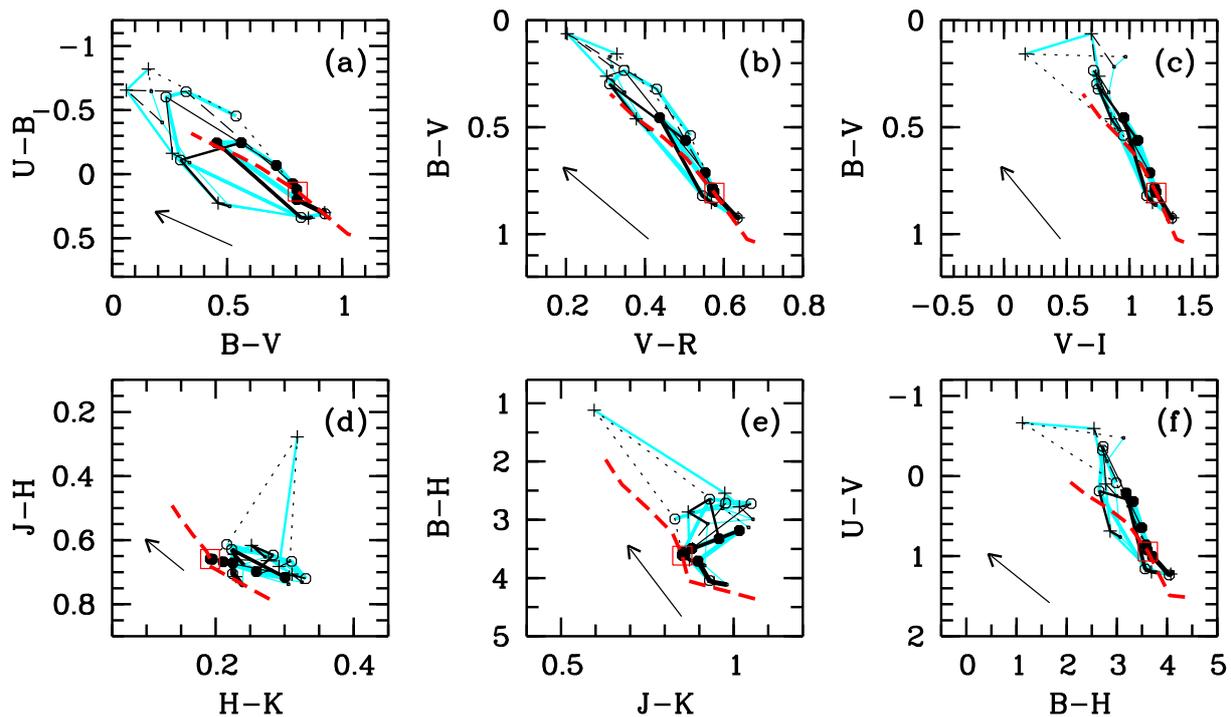}}
\caption{Variation in the position of a stellar system with the characteristic star-formation time 
scale $\tau_1 = 5$~Gyr (large empty square) on the two-color diagrams in the case when it 
experiences a secondary burst of star formation beginning 3, 1 Gyr, 300, 100, 30, and 10 Myr 
ago (black curves; the thickness of the solid curves decreases from $T_2 = 3$~Gyr to 
100~Myr, and the dashed and dotted curves correspond to $T_2 = 30$ and 10~Myr, 
respectively), with characteristic star-formation time scales $\tau_2 = 1$~Gyr (filled circles), 
100 (hollow circles), 10 (pluses), and 10~Myr (points) and with $M_2/M_1 = 1$. The systems 
with the same $\tau_2$ values are joined by cyan line segments whose thickness decreases with 
decreasing $\tau_2$. The thick red dashed curve shows the NCS for galaxies with an age of 
13~Gyr and zero initial metallicity. The remaining notation is as in Fig.~\ref{figure:fig2}.
\label{figure:fig4}}
\end{figure*}

Fig.~\ref{figure:fig3} presents the variations of the NCS in the two-color diagrams for galaxies with 
a secondary burst of star formation. The set of model sequences (96 curves) occupies an extended 
region in the diagrams, spreading to the right, left, and above the NCS, and substantially deviating 
from the absorption line. The largest deviations to the right and left of the NCS are observed in the 
$(U-B)\div(B-V)$ diagram (Fig.~\ref{figure:fig3}a).

A recent burst of star formation ($T_2 \sim 10-30$~Myr) shifts a stellar system toward the upper 
left corner of the two-color diagrams (an exception is the $(J-H)\div(H-K)$ diagram; 
Fig.~\ref{figure:fig3}d), even if this burst has a relatively modest power ($M_2/M_1 \sim 0.1$; 
Figs.~\ref{figure:fig3}a-\ref{figure:fig3}c, \ref{figure:fig3}e, \ref{figure:fig3}f). Weaker bursts 
($M_2/M_1 \sim 0.01$) shift a stellar system upward from the NCS in the $(B-V)\div(V-I)$ and 
$(U-V)\div(B-H)$ diagrams (Figs.~\ref{figure:fig3}c, \ref{figure:fig3}f). Thus, even a modest burst 
of star formation can appreciably decrease the $U-B$ and $B-V$ values, without influencing the 
longer-wavelength color indices.

Powerful bursts of star formation that occurred longer in the past ($T_2 \sim 3$~Gyr, 
$M_2/M_1 = 1$) can shift a stellar system downward and to the right along the NCS 
(Figs.~\ref{figure:fig3}a-\ref{figure:fig3}c, \ref{figure:fig3}e, \ref{figure:fig3}f). This is a 
consequence of the fact that the color indices of an old stellar population depend only weakly on 
its age, but grow substantially with increasing metallicity (Fig.~\ref{figure:fig2}). A comparison 
of the positions of the observed galaxies and model stellar systems with secondary bursts of 
star formation on the $(U-B)\div(B-V)$ diagram indicates that galaxies with the reddest $U-B$ colors 
are described well by models with secondary bursts of star formation occurring three or more Gyr 
ago (Figs.~\ref{figure:fig2}a, \ref{figure:fig3}a).

Stellar systems with secondary bursts of star formation occurring $\sim0.3-1$~Gyr ago are of 
special interest. Such systems are shifted to the left of the NCS in the $(U-B)\div(B-V)$, 
$(B-V)\div(V-R)$, $(B-V)\div(V-I)$, and $(U-V)\div(B-H)$ diagrams 
(Figs.~\ref{figure:fig3}a-\ref{figure:fig3}c, \ref{figure:fig3}f). For stellar systems with 
$\tau_1 \ge 5$~Gyr (galaxies with type Sa and later), this shift is perpendicular to the absorption 
line, and can in principle be detected.

Let us consider how the position of a stellar system varies in the two-color diagrams as a function 
of the parameters $\tau_2$ and $T_2$. Fig.~\ref{figure:fig4} presents the variations in the position 
of a stellar system with $\tau_1 = 5$~Gyr (a type Sa--Sb galaxy) in the case of a secondary burst 
of star formation with $M_2/M_1 = 1$.

Systems with recent bursts of star formation ($T_2 = 10-30$~Myr, $\tau_2 = 10$~Myr) with blue 
colors can be distinguished in all the two-color diagrams (with the exception of $H -K$; 
Fig.~\ref{figure:fig4}d). Models with such parameters describe star-forming regions in galaxies.

Models with bursts of star formation that ended in the more distant past ($T_2 = 3$~Gyr, 
$\tau_2 < 1$~Gyr) have redder colors than the original stellar systems (Fig.~\ref{figure:fig4}).

In the optical, for systems with the same characteristic decay times $\tau_2$, systems with 
smaller $T_2$ values have bluer colors. The dependence of the color on $\tau_2$ is more 
complex: when $\tau_2 < T_2$, systems with larger $\tau_2$ values have bluer colors, while 
systems with smaller $\tau_2$ values have bluer colors when $\tau_2 > T_2$. Thus, the 
bluest colors are observed for systems with $\tau_2 = T_2$ -- at the maximum SFR for the 
secondary burst (Fig.~\ref{figure:fig4}a-\ref{figure:fig4}c, \ref{figure:fig4}f).

\begin{figure*}
\vspace{8mm}
\centerline{\includegraphics[width=16cm]{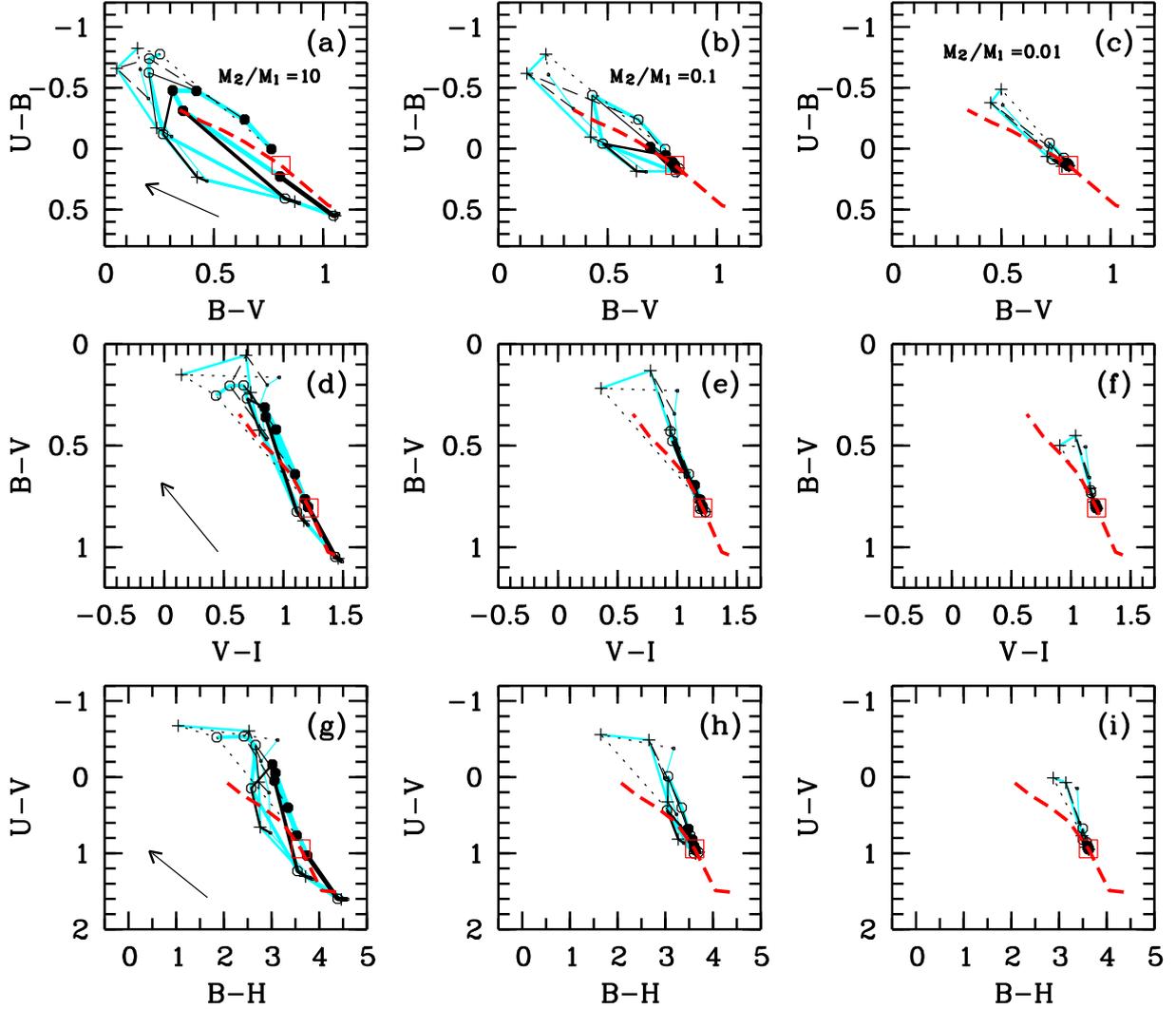}}
\caption{Variatons in the positions of stellar systems with a characteristic star-formation time 
$\tau_1 = 5$~Gyr with a secondary burst of star formation with $M_2/M_1 =$ {\bf (a, d, g)} 
10, {\bf (b, e, h)} 0.1, and {\bf (c, f, i)} 0.01 on the {\bf (a)--(c)} $(U-B)\div(B-V)$, 
{\bf (d)--(f)} $(B-V)\div(V-I)$, and {\bf (g)--(i)} $(U-V)\div(B-H)$ two-color diagrams. 
Notation is as in Fig.~\ref{figure:fig4}.
\label{figure:fig5}}
\end{figure*}

The parameters $\tau_2$ and $T_2$, their ratios, and their absolute values influence the various 
color indices differently. This is most clearly visible in the $(U-B)\div(B-V)$ diagram: systems 
with $\tau_2 < T_2$ are shifted to the left of the original positions without a secondary burst 
of star formation, the NCS, and the absorption line; systems with $\tau_2 > T_2$ are shifted to 
the right of the NCS and the absorption line (Fig.~\ref{figure:fig4}a). Similar behavior for the 
systems with secondary bursts of star formation is observed on the $(B-V)\div(V-R)$, 
$(B-V)\div(V-I)$, and $(U-V)\div(B-H)$ diagrams, although the deviations of these systems 
from the NCS and absorption line are smaller here than in the $(U-B)\div(B-V)$ diagram, and 
the systems themselves are shifted primarily to the right of the NCS (Figs.~\ref{figure:fig4}b, 
\ref{figure:fig4}c, \ref{figure:fig4}f).

In the $(J-H)\div(H-K)$ and $(B-H)\div(J-K)$ diagrams, secondary bursts with any parameters 
apart from $T_2 = 1$~Myr and $\tau_2 = 1$~Myr shift systems to the right of the NCS and 
the original position of the galaxy. Systems with prolonged secondary bursts ($T_2 = 1$~Gyr) 
are located systematically below systems with $T_2 \le 100$~Myr (i.e., they have higher $J-H$ 
and $B-H$ color indices; Figs.~\ref{figure:fig4}d, \ref{figure:fig4}e).

The positions of stellar systems with secondary bursts of star formation on the 
$(U-B)\div(B-V)$, $(B-V)\div(V-I)$, and $(U-V)\div(B-H)$ two-color diagrams as a function 
of the power of the burst ($M_2/M_1$) are shown in Fig.~\ref{figure:fig5}. Increasing the 
power $M_2/M_1 = 10$ only weakly influences the position of a stellar system in the diagrams. 
This is due to the fact that, with $M_2/M_1 \ge 1$, the main contribution to the optical radiation 
is made by relatively young stars formed in the second burst. For systems with $M_2/M_1 < 1$, 
the difference between the colors of the original galaxy and of a system with a secondary burst 
of star formation decrease systematically as the ratio $M_2/M_1$ is increased. The deviations 
from the position of the original galaxy and the NCS in the two-color diagrams are appreciable 
for systems with burst powers $M_2/M_1 = 0.1$, and can be detected if the burst was relatively 
recent ($T_2 \le 300$~Myr) and not prolonged ($\tau_2 \le 100$~Myr). Decreasing the burst 
power by another order of magnitude, to values $M_2/M_1 = 0.01$, means that only systems 
with recent secondary bursts with durations of the order of 10~Myr can be distinguished in the 
diagrams ($T_2 \le 30$~Myr, $\tau_2 \le 10$~Myr; Figs.~\ref{figure:fig5}c, \ref{figure:fig5}f, 
\ref{figure:fig5}i).

\begin{figure}
\vspace{6mm}
\centerline{\includegraphics[width=8.5cm]{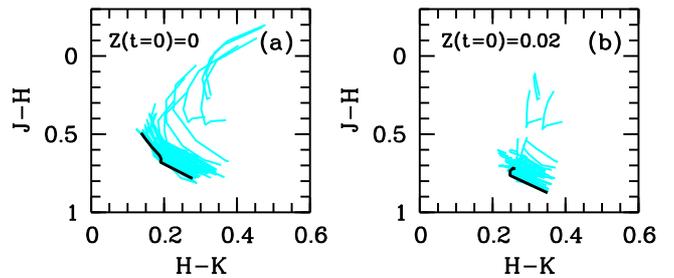}}
\caption{Variation of the NCS for galaxies with age 13~Gyr and a secondary burst of star 
formation having {\bf (a)} zero and {\bf (b)} solar initial metallicity (thick black curves) on the 
$(J-H)\div(H-K)$ two-color diagram. The remaining notation is as in Fig.~\ref{figure:fig3}.
\label{figure:fig6}}
\end{figure}

Since the IR diagram $(J-H)\div(H-K)$ serves as a good indicator of metallicity but reacts only 
weakly to the age of the stellar population, we used this diagram to estimate the variations of the 
NCS of galaxies with an age of 13~Gyr and various initial metallicities that have undergone 
secondary bursts of star formation (Fig.~\ref{figure:fig6}). Unfortunately, the cloud of model 
sequences for galaxies with zero initial metallicity cover the cloud of sequences for galaxies with 
the solar initial metallicity virtually completely. This reflects the fact that the ratio $Z$ for systems 
with different initial metallicities decreases as the systems evolve (in our case, from several 
orders of magnitude at time $t \approx 0$ to a factor of a few at time $t = 10-13$~Gyr).

\section{ANALYSIS OF THE OBSERVATIONAL DATA}

\subsection{Positions of Galaxy Components on Two-Color Diagrams}

Earlier, we determined the structural components for nine galaxies (nucleus, bulge, disk, spiral 
arms, bar, ring), obtained the color indices in the $UBVRI$ bands, and analyzed the positions 
of the components on two-color diagrams 
\citep{art99,art00,gus02a,gus02b,gus03a,gus03b,gus04,bru10}. In our current study, we 
measured the optical color indices for the components of the remaining 17 galaxies, and also 
determined the color indices of the structural components in the IR for nine of these galaxies 
(Table~\ref{table:tab2}).

\begin{figure*}
\vspace{8mm}
\centerline{\includegraphics[width=16cm]{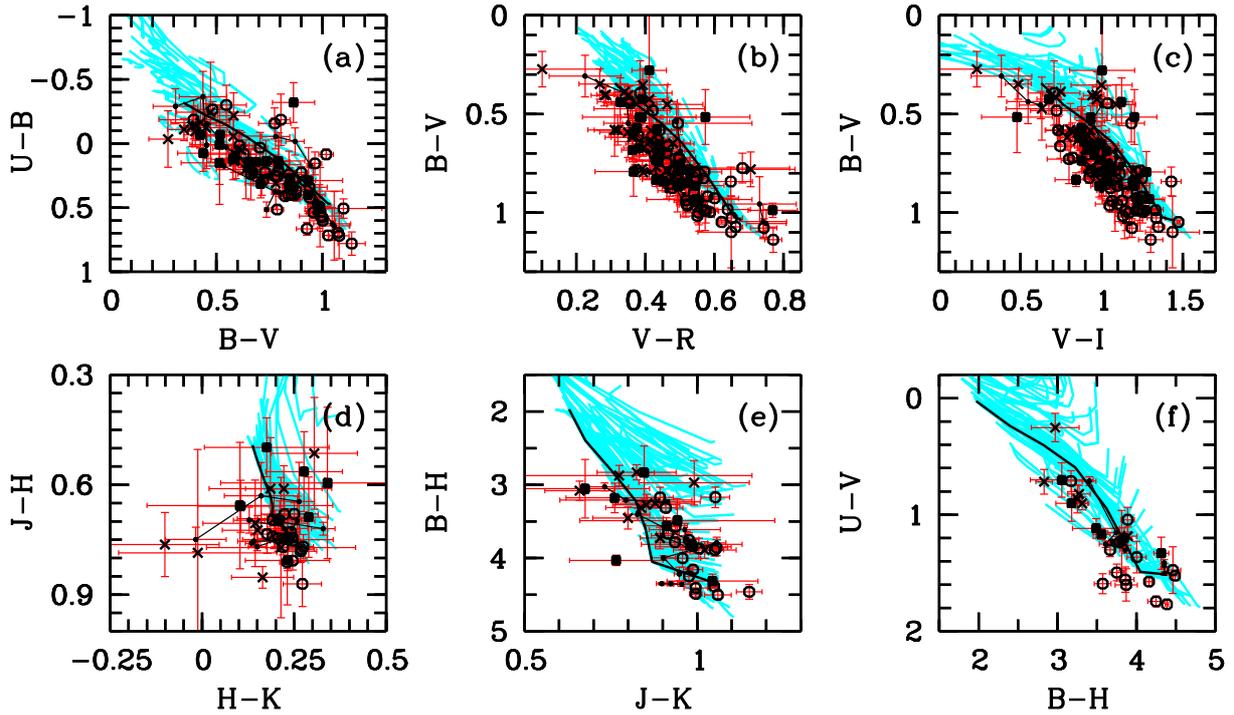}}
\caption{Positions of the nuclei and bulges (hollow circles), disks (filled squares), bars 
(points joined by line segments), and spiral arms and rings ($\times$'s) in the two-color diagrams. 
The red bars shown indicate the uncertainties. The remaining notation is as in 
Fig.~\ref{figure:fig3}.
\label{figure:fig7}}
\end{figure*}

Fig.~\ref{figure:fig7} shows the positions of the galactic components (nuclei, bulges, disks, 
spiral arms, bars, rings) on the two-color diagrams. We chose areas outside regions of star 
formation to determine the color indices of spiral arms, rings, and bars. Overall, the color 
indices of these galactic components agree with the model NCS within the uncertainties. 
However, the clouds of point indicating the positions of components in the optical diagrams 
are shifted to the left of the model NCS (Fig.~\ref{figure:fig7}a-\ref{figure:fig7}c, 
\ref{figure:fig7}f), and to the right in the $(B-H)\div(J-K)$ diagram (Fig.~\ref{figure:fig7}e). 
This shift cannot be explained by a failure to correctly take into account absorpton in the 
galaxies, since this would shift the positions of the components along the NCS, or by an effect 
of the age of the galaxies (Fig.~\ref{figure:fig2}). Our data agree well with the integrated color 
indices of galaxies: the cloud of points indicating the positions of galaxies on the optical 
two-color diagrams is also systematically shifted to the left from the model NCS. The 
$(B-V)\div(V-I)$ diagram illustrates this especially well (compare Figs.~\ref{figure:fig2}c and 
\ref{figure:fig7}c). Thus, the observed colors of the galaxies and their components are, on 
average, slightly redder than the model colors. We will discuss possible origins of this deviation 
in detail in the next section. Note that the uncertainties in the characteristic SFR decay time 
$\tau_1$ due to the deviations of the positions of the galaxy components from the NCS do not 
exceed the uncertainties associated with uncertainties in the color indices.

The values of $\tau_1$ derived from $U-B$ are systematically lower than the $\tau_1$ values 
derived from $B-V$ by a factor of 1.5, while the $\tau_1$ values derived from $V-R$ and 
$V-I$ are systematically higher by a factor of $1.8-2$. Thus, the absolute magnitude of the 
characteristic decay time for the SFR can be determined from the positions of components on 
the two-color diagrams to within about a factor of three. The relative values of $\tau_1$ for the 
components of individual galaxies can be determined more precisely, and depend only on the 
uncertainty in the color indices and the difference of the internal absorption in a structural 
component from the mean absorption for the galaxy as a whole.

\begin{figure}
\vspace{6mm}
\centerline{\includegraphics[width=8.5cm]{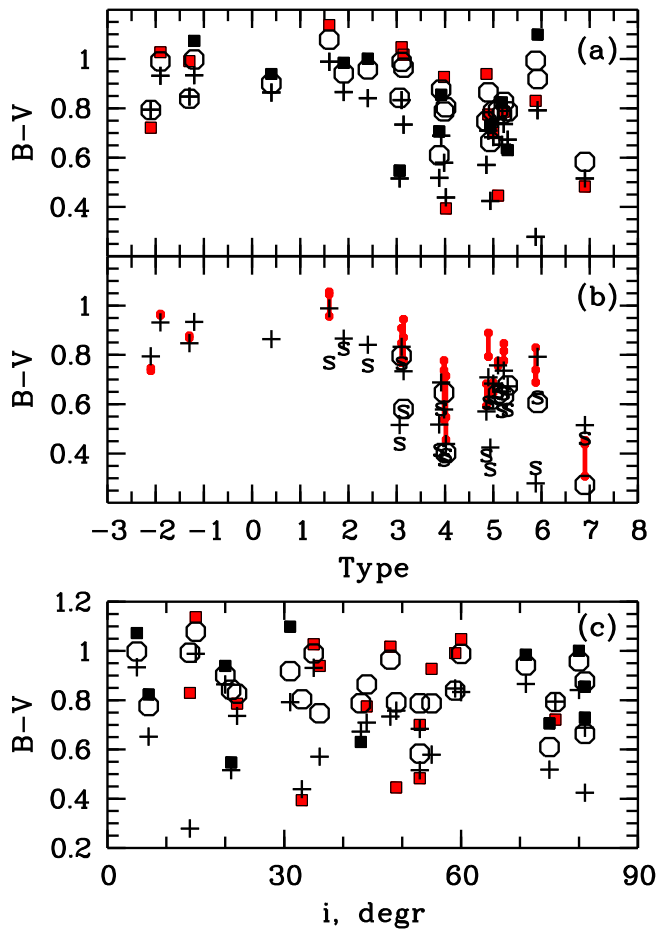}}
\caption{Dependence of the $B-V$ color index on themorphological type of the galaxy: {\bf (a)} 
for nuclei of galaxies with bars (red squares), nuclei in galaxies without bars (black squares), 
bulges (circles), and disks ($\times$'s); {\bf (b)} for disks ($\times$'s), spiral arms (''s''), rings 
(circles), and bars (points joined by red line segments). {\bf (c)} Dependence of $B-V$ on the 
inclination of the galaxy for nuclei, bulges, and disks (same notation as in panel {\bf (a)}).
\label{figure:fig8}}
\end{figure}

The locations of the vast majority of the galactic components, including spiral arms and rings, 
along the NCS in the two-color diagrams makes it possible to crudely estimate their 
photometric characteristic ages (as a function of $\tau_1$, assuming that the absorption in 
the component does not differ strongly from the mean vaue calculated for the inclination of the 
galactic disk), but not the time and power of the secondary burst of star formation. An 
exception is extreme cases, for example, the nuclei with active star formation in the galaxies 
NGC~245 and NGC~3726, which are shifted to the right of the NCS toward the region of 
models with secondary bursts of star formation (Fig.~\ref{figure:fig7}a, \ref{figure:fig7}c). 
As expected, galactic nuclei and bulges lie in the lower part of the NCS (i.e., they have redder 
colors), while, as a rule, disks, spiral arms, and rings are located in the central and upper parts 
of the NCS (Figs.~\ref{figure:fig7}a-\ref{figure:fig7}c, \ref{figure:fig8}a, \ref{figure:fig8}b). 
The structural components are essentially not observed in the upper part of the NCS in the IR 
diagrams (Figs.~\ref{figure:fig7}d-\ref{figure:fig7}f). This is due to a selection effect: the 
fraction of lenticular galaxies observed in the IR is much higher than the average for the sample 
(Tables~\ref{table:tab1} and \ref{table:tab2}).

Figs.~\ref{figure:fig2}a-\ref{figure:fig2}c and \ref{figure:fig7}a-\ref{figure:fig7}c show that the 
cloud of points characterizing the color indices of the galaxies and their components fully cover 
the length of the NCS, and include appreciable shifts along the NCS. The color indices of the 
components do not depend on the inclination of the galactic disks (Fig.~\ref{figure:fig8}c). This 
indicates that the internal absorption of the galaxies as a function of the disk inclination is 
satisfactorily described by the true absorption in the galaxies, and can be used to make statistical 
estimates for large samples of galaxies and their components.

Figs.~\ref{figure:fig8}a and \ref{figure:fig8}b show the $B-V$ color indices for various 
components along the Hubble sequence of galaxies. The numerical morphological types of the 
galaxies were taken from Table~\ref{table:tab1}; for galaxies of the same morphological type, their 
order along the horizontal axis corresponds to their position in Table~\ref{table:tab1}.

The galaxies clearly divide into two groups according to their photometric characteristics: the 
S0--Sb galaxies (Types~$\le3$) and the Sb--Sd galaxies (Types~$\ge3$). All the components in 
galaxies in the first group, including disks, spirals, and rings, have red colors ($B-V >0.7$). 
The lenticular galaxies and early-type spiral galaxies (to Sab inclusive) cannot be distinguished 
based on their photometric characteristics. The only exception is the photometric characteristics 
of bars: the colors of the bars of S0 galaxies do not change with distance from the center, 
while the bars in early-type spiral galaxies become bluer toward their ends (Fig.~\ref{figure:fig8}b). 
All the components in galaxies of the second group (Sb--Sd) are systematically bluer than the 
corresponding components of galaxies of the first group. This difference is less pronounced for 
the bulges and bars, but is clearly visible for the disks and spiral arms (Figs.~\ref{figure:fig8}a, 
\ref{figure:fig8}b). Within the second group, on average, the galaxies and their components become 
bluer along the Hubble sequnce, although the correlation between the type of galaxies and the color 
of its components is weak. The Sb galaxies include galaxies in both the first group (NGC~5351) 
and the second group (NGC~245 and IC~1525; Figs.~\ref{figure:fig8}a, \ref{figure:fig8}b).

The $B-V$ color indices of all the galactic components lie in the range from 0.25 to 1.15. 
Components with $B-V<0.5$ deviate to the right or left of the NCS in the two-color diagrams, 
toward the region of models with secondary bursts of star formation 
(Figs.~\ref{figure:fig7}a-\ref{figure:fig7}c). All types of structural components except for bulges 
are observed among such objects (Figs.~\ref{figure:fig8}a, \ref{figure:fig8}b). The reddest 
components with $B-V>1.1$ could also have a complex star-formation history: they occupy 
regions corresponding to old, but chemically enriched, stellar systems in the two-color diagrams.

\subsection{Nuclei, Bulges, and Disks}

The reddest components are the bulges (Fig.~\ref{figure:fig8}a), whose $B-V$ color indices 
range from 0.8 to 1.05 (corresponding to $\tau_1 < 5$~Gyr) for S0--Sb galaxies and from 0.6 to 
1.0 ($\tau_1 = 1-7$~Gyr) for later type galaxies. The vast majority of the bulges lie along the NCS, 
and are described well bymodels with exponentially decaying SFRs with $\tau_1 \approx 1-5$~Gyr. 
Exceptions are the bulge in NGC~6217, which deviates to the right of the NCS in the 
$(U-B)\div(B-V)$ diagram, the bulge in NGC~245, which deviates to the right of the NCS in the 
$(B-V)\div(V-R)$ and $(B-V)\div(V-I)$ diagrams, and the bulge in NGC~266, which lies below 
the NCS.

The nuclei of most of the studied galaxies are slightly redder than their bulges (by $\approx 0.05$ 
in $B-V$; Fig.~\ref{figure:fig8}a). However, roughly one-quarter of the galaxies have relatively 
blue nuclei, with the difference in $B-V$ for the nucleus and bulge sometimes exceeding 0.3. Due 
to this, the range of $B-V$ values for the nuclei is larger than for the bulges: for nuclei of galaxies 
in the first group, $B-V = 0.75-1.15$, while nuclei of Sb--Sd galaxies have $B-V = 0.4-1.1$ 
(Fig.~\ref{figure:fig8}a). We note especially the four Sb--Sd galaxies NGC~245, NGC~6217, 
NGC~3726, and NGC~5585 with blue nuclei with $B-V<0.6$ (which corresponds to 
$\tau_1 \sim 10-13$~Gyr; Fig.~\ref{figure:fig8}a). Three of these galaxies have bars, and the 
fourth -- NGC~245 -- is a Makarian galaxy with an asymmetrical spiral structure. The only S0 
galaxy with a relatively blue nucleus is NGC~7351, which also has a bar. The blue nuclei lie in the 
upper part of the NCS, although the nuclei of NGC~245 and NGC~3726 are shifted to the right 
of the NCS in the $(B-V)\div(V-I)$ diagram (Fig.~\ref{figure:fig7}c). 
Deviations of nuclei to the right of the NCS are also observed for NGC~7678 in the 
$(U-B)\div(B-V)$ and $(B-V)\div(V-R)$ diagrams, for IC~1525 in the $(U-B)\div(B-V)$ diagram, 
and for NGC~783 in the $(J-K)\div(B-H)$ diagram (Figs.~\ref{figure:fig7}a, \ref{figure:fig7}b, 
\ref{figure:fig7}e).

The nucleus of NGC~266 (SBbc, inclination $15\degr$) is the reddest component in the studied 
galaxies. It has the highest $U-B$, $B-V$, and $V-R$ values, and is located below the NCS in the 
two-color diagrams, in the region of models with a secondary burst of star formation in the distant 
past (more than three billion years ago). The nucleus of NGC~266 is shifted to the right of the 
NCS in the $(B-V)\div(V-R)$ diagram (Fig.~\ref{figure:fig7}b).

The color indices of the disks are always lower than those of the spherical components (nuclei 
and bulges) of their galaxies, with this difference increasing along the Hubble sequence 
(Fig.~\ref{figure:fig8}a). An exception is galaxies with blue nuclei, in which, as a rule, the color 
characteristics of the nuclei and disks are very similar. The $B-V$ values for the disks of early-type 
galaxies are 0.8--1.0, while those for Sb--Sd galaxies are 0.3--0.8 (Fig.~\ref{figure:fig8}a). The 
bluest disks ($B-V \le 0.5$) are those in late-type galaxies with extremely blue nuclei (NGC~245, 
NGC~6217, NGC~3726, and NGC~5585) and the disk in the Sc galaxy NGC~3184, which has 
a bar and a moderately blue nucleus (Fig.~\ref{figure:fig8}a). The disks of NGC~245, NGC~3184, 
and NGC~6217 are shifted to the right of the NCS on the $(B-V)\div(V-R)$ and/or $(B-V)\div(V-I)$ 
diagrams (Figs.~\ref{figure:fig7}b, \ref{figure:fig7}c).

The disk of the lenticular galaxy NGC~6340 occupies an anomalous position in the upper right part 
of the $(U-B)\div(B-V)$ diagram, far from the NCS (Fig.~\ref{figure:fig7}a). This galaxy was studied 
in detail by \citet{chi09}, who showed that NGC~6340 has a fairly complex structure and evolution; 
in particular, it has undergone mergers and absorptions in the past.

Unfortunately, the IR color indices do not provide additional information about the stellar population 
of the galaxy components, due to their large uncertainties (Figs.~\ref{figure:fig7}d-\ref{figure:fig7}f).

\subsection{Spiral Arms, Rings, and Bars}

\begin{figure*}
\vspace{8mm}
\centerline{\includegraphics[width=16cm]{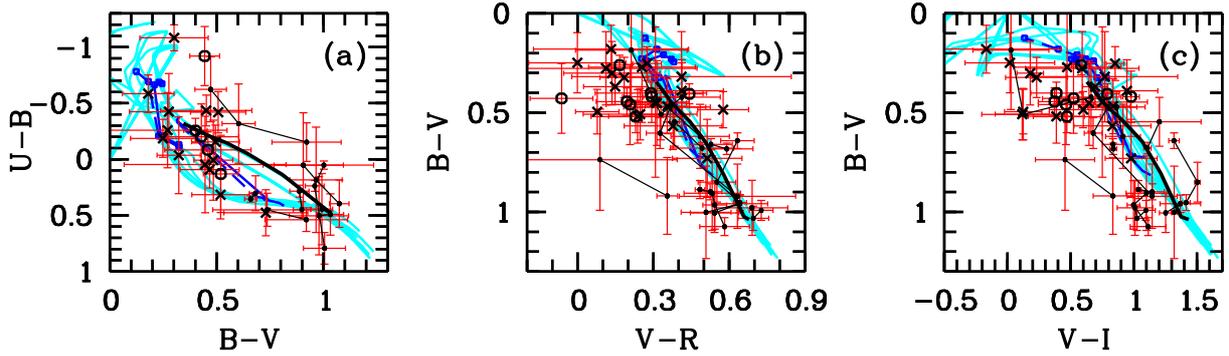}}
\caption{Positions of spiral arms ($\times$'s), rings (hollow circles), and bars (points joined by line 
segments) on the optical two-color diagrams, after subtracting the contributions of the disk and bulge 
from the luminosity. The measurement uncertainties are shown by red bars. The cyan curves show 
evolutionary tracks for stellar systems with initial metallicity $Z(t = 0) = 0.008$ and 0.02 and 
characteristic star-formation decay time scales $\tau = 1$, 10, 100 Myr, 1, and 3 Gyr. The blue thin 
curves show isochrones for ages of $t = 100$~Myr (points) and 1~Gyr for systems with 
$Z(t = 0) = 0.008$ (dotted curves) and 0.02 (solid curves). The black thick curves show the NCS for 
the galaxies.
\label{figure:fig9}}
\end{figure*}

\begin{figure}
\vspace{6mm}
\centerline{\includegraphics[width=8.5cm]{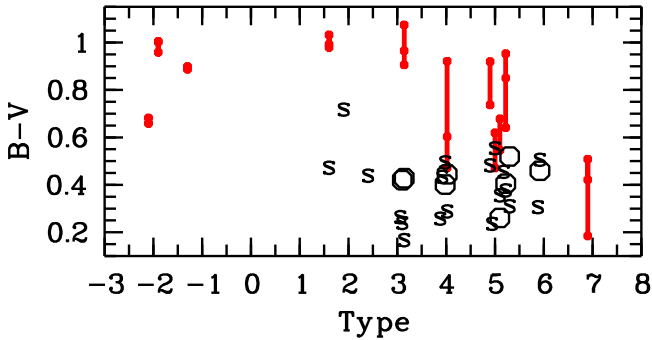}}
\caption{Same as Fig.~\ref{figure:fig8}b for the spiral arms, rings, and bars, after subtraction 
of the contributions of the disk and bulge.
\label{figure:fig10}}
\end{figure}

As a rule, spiral arms and rings are slightly bluer than the surrounding disk. The difference between the 
$B-V$ values for spiral arms and rings does not exceed 0.1 for all the galaxies except for NGC~2336 
and NGC~5585 (Fig.~\ref{figure:fig8}b). NGC~5585 is a low-surface brightness Sd galaxy, and has 
an anomalously blue ring. The $B-V$, $V-R$, and $V-I$ values of the ring are the lowest among all 
the components of the galaxies studied (Figs.~\ref{figure:fig7}b, \ref{figure:fig7}c). We studied this 
galaxy in \citet{bru10}, where we showed that the center of the ring does not coincide with the center 
of the galaxy. This structure of NGC~5585 could be understood as the result of a recent external 
interaction (merger) \citep{bru10}. The NGC~5585 ring lies above and to the left of the NCS in the 
two-color diagrams, in the region of models with a secondary burst of star formation 
(Figs.~\ref{figure:fig7}a-\ref{figure:fig7}c).

The spiral arms (and rings, when present) of NGC~266, NGC~2336, NGC~5605, NGC~6946, and 
IC~1525 are appreciably bluer than their surrounding disks (Fig.~\ref{figure:fig8}b). These galaxies 
possess well developed, bright spiral arms that are clearly defined against the background of the disk, 
even in longwavelength bands.

The $B-V$ values of the spiral arms and rings of Sa--Sb galaxies lie within the narrow range from 0.75 
to 0.85, while $B-V$ for the spiral arms and rings of galaxies in the second group are 0.25--0.7. The 
bluest spiral arms (rings), with $B-V<0.5$, deviate to the right or left of the NCS on the 
$(B-V)\div(V-R)$ and $(B-V)\div(V-I)$ diagrams (Figs.~\ref{figure:fig7}b, \ref{figure:fig7}c, 
\ref{figure:fig8}b). Such spiral arms are observed in eight galaxies in our sample: five with blue disks 
(see above) and three Sbc--Scd galaxies with blue spirals against the background of relatively red 
disks (NGC~2336, NGC~5605, and NGC~6946).

In terms of their photometric characteristics, the bars occupy an intermediate position between the 
nuclei (bulges) and the spiral arms. As a rule, the inner regions of bars have color indices close to 
those of the nucleus or bulge (in galaxies with blue nuclei). The outer regions of bars have colors 
close to those of the disks (in S0 galaxies) or spiral arms (Figs.~\ref{figure:fig8}a, \ref{figure:fig8}b). 
They occupy regions along the lower and middle part of the NCS on the two-color diagrams. One 
exception is NGC~5585, which has a complex evolution, whose very blue bar is located in the upper 
part of the NCS (Figs.~\ref{figure:fig7}a-\ref{figure:fig7}c). The bars of IC~1525 and of NGC~266 
are shifted to the right of the NCS on the $(U-B)\div(B-V)$ and $(B-V)\div(V-R)$ diagrams, 
respectively.

\begin{figure*}
\vspace{8mm}
\centerline{\includegraphics[width=16cm]{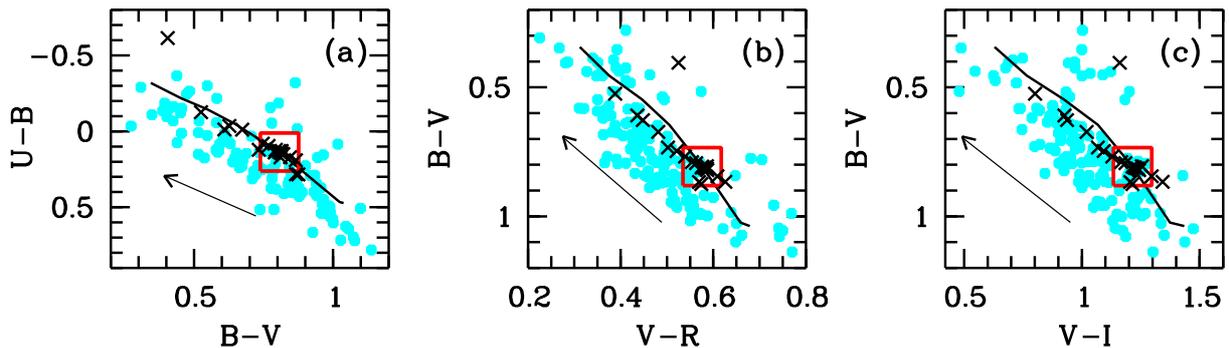}}
\caption{Variations in the positions of stellar systems with characteristic star-forming time scale 
$\tau_1 = 5$~Gyr ($\times$'s) on the two-color diagrams for various model parameters, relative 
to systems with the model parameters used in this study (red square). The black curve shows 
the NCS and the filled cyan circles the positions of the galaxy components 
(Fig.~\ref{figure:fig7}). The arrows indicate the variation of the color indices upon correction 
for absorption equal to $A(V) = 1.0^m$.
\label{figure:fig11}}
\end{figure*}

Since the spirals, bars, and rings develop against the background of their associated disks, we also 
determined the brightnesses and color indices after subtracting the contribution of the disk (as well 
as the bulge, for inner regions of the bars) from the total luminosity in the corresponding photometric 
bands. Unfortunately, it was not possible to determine the color indices of some of the spiral arms, 
bars, and rings that were only weakly distinguished against the disk after subtracting the disk 
radiation, due to their large uncertainties. The uncertainties in the IR were too large to consider the 
galaxy components in the IR diagrams.

Fig.~\ref{figure:fig9} shows the positions of the spiral arms, bars, and rings after subtracting the 
contributions of the disk and bulge. In contrast to the galaxy components presented in 
Fig.~\ref{figure:fig7}, the spiral arms, rings, and bars are not grouped along the NCS after the 
subtraction of the disk contribution. The points characterizing the color indices of the spiral arms 
and rings are located at the upper end of the NCS, and these color indices are contained within 
fairly narrow ranges: $-0.6 < U-B < 0.1$, $0.2 < B-V < 0.5$, $0.0 < V-R < 0.4$ 
(Figs.~\ref{figure:fig9}a, \ref{figure:fig9}b). The bars can be clearly distinguished from the spiral 
arms and rings in the diagrams, and are located near the lower part of the NCS. Note the absence 
of any correlation between the color characteristics of the spiral arms and rings and the 
morphological types of the galaxies (Fig.~\ref{figure:fig10}).

Unfortunately, the uncertainties in the color indices for the components, which reach $\pm0.2$ 
for $U-B$, $B-V$, and $V-R$ and $\pm0.3$ for $V-I$, hinder accurate determination of the 
component ages; only qualitative crude estimates are possible. The galaxy components are 
arranged along the evolutionary tracks of stellar systems with roughly solar metallicity in the 
two-color diagrams. The best indicator of this is the $(U-B)\div(B-V)$ diagram 
(Fig.~\ref{figure:fig9}a). Most of the spiral arms are located in the region of models with ages 
of 100~Myr -- 1~Gyr and characteristic star-formation decay times $\tau \le 1$~Gyr; the rings 
are somewhat older ($t \sim 1$~Gyr). The color indices for most of the bars correspond to 
stellar systems with ages of several Gyr (Fig.~\ref{figure:fig9}a).

The $(B-V)\div(V-R)$ and $(B-V)\div(V-I)$ diagrams cannot be used to estimate the 
component ages: the isochrones of stellar systems with different $\tau$ values are located 
along the evolutionay tracks (Fig.~\ref{figure:fig9}b, \ref{figure:fig9}c). However, the 
$(B-V)\div(V-I)$ diagram can serve as a good indicator for the ages of the youngest 
components, with ages less than 100~Myr. Systems with ages of less than 100~Myr have 
$V-I \approx 0.0$ and are located above and to the left of the NCS (Fig.~\ref{figure:fig9}c).

The bluest components in the sample are the spiral arms of NGC~2336, NGC~6217, 
NGC~7721, and IC~1525 (these are all giant galaxies with numerous star-forming regions), 
the rings of NGC~3726 and NGC~6217, and the outer part of the bar of the peculiar Sd 
galaxy NGC~5585 (the stellar populations of these components have ages of 10--100~Myr; 
Figs.~\ref{figure:fig9}a-\ref{figure:fig9}c).

The spiral arms of NGC~532 are anomalously red. This can be explained by the high 
inclination of the galaxy ($i = 71\degr$) and the presence of a powerful dust disk in the 
galaxy \citep{gus06a}. It is likely that the absorption in the spiral arms of NGC~532 is 
appreciably higher than the mean for this galaxy.

The color indices of the bars after subtraction of the contributions of the disk and bulge are 
often higher than those before this subtraction (compare Figs.~\ref{figure:fig8}b and 
\ref{figure:fig10}). This is especially true of the inner regions of the bars. This may indicate 
that the stars in the inner part of the bars are older than the stars in the surrounding disk, or 
that the inner regions of the bars contain a large concentration of dust.

\section{DISCUSSION}

\subsection{The Model NCS and the Observed Color Indices of Galaxy Components}

As is shown by Figs.~\ref{figure:fig2}a-\ref{figure:fig2}c, \ref{figure:fig7}a-\ref{figure:fig7}c, 
the observed color indices of the galaxies and their components are systematically redder than 
the model color indices. The points representing the former are shifted to the left and 
downward by $\sim0.1$ in each color index relative to the model NCS in the optical two-color 
diagrams. Varying the input parameters of the models of \citet{fio97} enabled us to consider 
possible origins of this deviation: differences of the time dependence of the SFR from the 
model dependence, differences between the real IMF and the Salpeter IMF with the specified 
upper and lower limits, complex structure of the stellar system that differs from a disk or 
spheroidal geometry, differences of the physical parameters of the stars and the interstellar 
medium from the model values.

In particular, we considered stellar systems with a Salpeter IMF ($\alpha = -2.35$) with upper 
mass limits from $30M\sun$ to $120M\sun$, and also IMFs with shallower ($\alpha = -1.35$) 
and steeper ($\alpha = -3.35$) profiles. We obtained the color characteristics of stellar systems 
for the classical model of \citet{san86} with an exponential decay of the SFR with 
${\rm SFR}(t) = 1/\tau \exp(-t/\tau)$ and also with a modified Sandage model (see Eq.~(1)) with 
coefficients ${\rm SFR}(t)/{\rm SFR}_1(t, \tau_1) = 0.1, 0.3, 3$, and 10.

Variation of the IMF, the dependence SFR($t$), and other parameters either influence the color 
indices of stellar systems with zero initial metallicity and age $\sim10-13$~Gyr only weakly, or 
shift the positions of the model stellar systems to the left and upward along the NCS 
(Fig.~\ref{figure:fig11}). The deviation of systems with model parameters different from those 
used in our study from the NCS do not exceed 0.1 in $B-V$ and $V-I$ and 0.05 in $U-B$ and 
$V-R$ (the only exception is the model with the shallow IMF with $\alpha = -1.35$, which has 
very blue $U-B$ and $B-V$ color indices; Fig.~\ref{figure:fig11}). These deviations are too 
small to explain the differences between the observed and model photometric parameters of the 
galaxies and their components. Moreover, a shift of the color sequences to the left and upward 
on the diagrams (for example, for the models with ${\rm SFR}(t)=0.1{\rm SFR}_1(t, \tau_1)$, 
which have the smallest $V-R$ and $V-I$ colors for the given $\tau_1$) leads to a large shift 
between the observed and model colors along the absorption line, which cannot be explained 
using modern estimates of the internal absorption in the galaxies due to the inclinations of their 
disks.

Thus, variations in the input model parameters cannot explain the differences between the 
observed and model photometric parameters of the galaxies and their components. We suggest 
that these differences may be due to differences in the zero points for the $UBVRI$ photometric 
system used in the PEGASE.2 program from the standard values. The problem of absolute 
calibration in the standard Johnson--Cousins $UBVRI$ photometric system remains a subject 
of discussion \citep{bes12}. This is due to differences in the photometric parameters of stars 
that are not on the main sequence in different stellar libraries, as well as uncertainties associated 
with the shapes of the transmission curves for the filters and the magnitudes and color indices 
of Vega (the zero points in the bands of this photometric system).

\subsection{Estimates of the Ages of the Stellar Populations and the Star-Formation History in 
the Galaxy Components based on Multi-color Photometry}

We showed in Section~4.1 that, on average, the internal absorption of the galaxies determined 
by the inclination of their disks can be adequately described by the real absorption in the galaxy, 
and can be used to study the photometric parameters of galaxies and their components for large 
samples of objects. In particular, this enables studies of variations of the photometric 
characteristics of the components along the Hubble sequence of galaxies.

The color characteristics of the components in early- and late-type galaxies differ appreciably 
(Fig.~\ref{figure:fig8}a, \ref{figure:fig8}b). Sb galaxies are the least uniform in terms of their 
photometric parameters. Some galaxies of this type have photometric properties corresponding 
to early-type galaxies, while the properties of others correspond to late-type galaxies. The 
morphological type Sb is sometimes assigned to intermediate-type galaxies 
\citep[see, e.g.,][]{mos10}. The characteristics of the disks (colors, relative thickness, 
mass-to-light ratio at the center) and bulges (Sersic parameter, ratio of the bulge to the disk 
luminosity) of Sb galaxies also reflect the inhomogeneity of the objects assigned to this 
morphologicl type \citep{gri98,mol04,gus07}. It is striking that, in spite of the appreciable 
differences in morphology, the components of lenticular galaxies exhibit essentially the same 
photometric characteristics as those of Sa galaxies.

The colors of the galactic nuclei correspond to those of stellar systems with old populations 
with ages $>7$~Gyr, and depend only weakly on the morphological types of the galaxies. The 
presence of a bar has a larger influence on the color characteristics (the mean age of the stellar 
population) of the nuclei. Radial gas motions along the bar can stimulate star formation in the 
nucleus over a prolonged period \citep{ars89,fri95}.

Unfortunately, the uncertainties in the color indices of the galaxy components are not sufficiently 
low to enable reliable determination of the compositions of their stellar populations and their 
star formation histories (compare Figs.~\ref{figure:fig3} and \ref{figure:fig7}). Another problem 
is that the absorption in individual galaxies or galaxy components may not correspond to the 
absorption calculated for the inclination of the galactic disks. This different could be substantial 
for the irregular components (spiral arms, rings, bars). Moreover, the distribution of dust, and 
as a consequence the amount of selective absorption, could be different in different spiral arms 
in the same galaxy. NGC~628 provides an example, where a dust lane passes through the inner 
edge of one spiral arm, and through the center of the stellar spiral of the opposite arm 
\citep{gus13}.

The color indices of the spiral arms correspond to those of stellar systems with ages 
$\sim1$~Gyr (Figs.~\ref{figure:fig9}a, \ref{figure:fig9}c). The age of the stellar population in 
the spiral arms is not the same at different galactocentric distances. In the vicinity of the 
corotation radius, the rotational velocity of the spiral pattern is equal to the rotational velocity 
of the disk, while the difference between these velocities increases with distance from the 
corotation radius. Thus, the maximum age of the stellar population in the spiral arms should 
be observed at distances from the galactic center corresponding to the corotation radius. In 
this region of the spiral arms, and also in rings, the characteristic time for a star to leave a given 
component is determined by the radial velocity dispersion for young stars.

In most galaxies, there is no secondary burst of star formation, or it is weak. The deviations 
of the components in these galaxies to the right or left of the NCS in the two-color diagrams 
are within the uncertainties in their color indices. However, the positions of the components of 
some galaxies in the two-color diagrams cannot be explained in the framework of evolutionary 
models with an exponentially decaying star-formation rate, as described by Eq.~(1). These 
galaxies have apparently undergone external interactions (mergers) in the past, which have 
stimulated secondary bursts of star formation in them. Many of these galaxies display signs 
of morphological peculiarities. We consider estimates of the parameters of the secondary burst 
of star formation and the compositions of the stellar populations in such galaxies below.

\subsection{Compositions of the Stellar Populations in the Components of Some Galaxies}

{\bf NGC 245.} This galaxy has a blue nucleus with a nearly constant star-formation rate over 
the past several billion years (the characteristic decay time for the SFR is 
$\tau_1 \approx 10$~Gyr). NGC~245 is a Makarian galaxy with asymmetric spiral arms and 
numerous star-forming complexes. The color indices of the disk and nucleus coincide within 
the uncertainties. According to its photometric characteristics, this is a late-type Sb galaxy. 
The components of NGC~245 lie to the left of the NCS on the $(U-B)\div(B-V)$ diagram and 
to the right of the NCS on the $(B-V)\div(V-I)$ diagram. This can be explained in models in 
which a secondary burst of star formation has ended ($\tau_2 < T_2$). The disk is described 
well by a model with a moderately powerful ($M_2/M_1 = 0.1$) burst of star formation 
occurring about 300~Myr ago; the bulge of NGC~245 is described well by a model with a 
powerful secondary burst of star formation long in the past ($T_2 \sim 3-10$~Gyr).

{\bf NGC 266.} This classical, symmetrical, early-type barred galaxy is viewed nearly face-on. 
It is distinguished by its anomalously red colors. The $U-B$, $B-V$, $V-R$, and $V-I$ values 
for the disk correspond to the integrated color indices of elliptical galaxies, and the color 
indices of the nucleus and bulge cannot be described using models with an exponential decay 
in the SFR and zero initial metallicity. All the components of NGC~266 -- the nucleus, bulge, 
disk, bar -- have the highest $B-V$ values for these components among our sample galaxies 
(Figs.~\ref{figure:fig8}a, \ref{figure:fig8}b). The relative positions of the components of 
NGC~266 in the two-color diagrams are typical for galaxies. The photometric characteristics 
of the components of NGC~266 are described well by models with a standard star-formation 
history in an initially chemically enriched stellar system (Fig.~\ref{figure:fig2}; the NCS for 
systems with $Z(t = 0) = Z\sun$). These colors can be explained by a secondary burst of star 
formation in the initial stage of the galaxy's evolution ($T_2 \sim 10$~Gyr), for example, as a 
result of the merger of two stellar systems. An alternative explanation of the anomalous 
photometric parameters of NGC~266, namely, the presence of a relatively large amount of dust, 
seems less probable to us; an excess of dust in a galaxy can be explained only as the result 
of non-standard evolution of the stellar system.

{\bf NGC 3184.} This is a late-type galaxy with numerous star-forming regions \citep{hod82} 
and a high integrated SFR \citep{you96,mar97}. The outer regions of the spiral arms have a 
curved, asymmetrical shape. The disk is located above and to the right of the NCS on the 
$(B-V)\div(V-R)$ and $(B-V)\div(V-I)$ diagrams. The photometric parameters of the 
NGC~3184 disk are characteristic of spiral arms. Apparently, active star formation is occurring 
in the NGC~3184 disk, which can be described in a model with numerous brief 
($\sim10-100$~Myr) bursts of star formation. This is also indicated by the high, diffuse 
background in the H$\alpha$ line in the disk between the spiral arms of the galaxy 
\citep{gus02b}. Clarifying the star-formation history in the disk of NGC~3184 will require 
surface photometry in $U$ and other UV bands.

{\bf NGC 3726.} This is a late-type galaxy with a bar, ring, and flocculent spiral arms. The outer 
regions of NGC~3726 have a well defined asymmetric shape 
\citep[which was not studied by us in][]{gus02a}. The positions of the components on the 
two-color diagrams are typical for Sc galaxies. An exception is the nucleus, which is the bluest 
component of the galaxy and deviates to the right of the NCS on the $(B-V)\div(V-I)$ diagram. 
The position of the nucleus of NGC~3726 on the two-color diagrams corresponds best to 
models with a moderate ($M_2/M_1 = 0.1$), relatively recent secondary burst of star formation 
($T_2 \sim 300$~Myr and $\tau_2 \sim 100$~Myr). Unfortunately, we did not observe this 
galaxy in the $U$ band, hindering unambiguous determination of the parameters of the nuclear 
burst of star formation.

{\bf NGC 5585.} This is a very late-type small galaxy with low surface brightness. It has a 
pronounced asymmetry: the centers of the galaxy, of the bar, and of the ring do not coincide 
\citep{bru10}. It is the bluest galaxy of our sample. The SFR in NGC~5585 is essentially 
constant: the positions of the components in the two-color diagrams correspond to 
$\tau_1 = 7-20$~Gyr. The colors of the ring and outer parts of the bar suggest that the ages 
of their stellar populations do not exceed 100~Myr.

{\bf NGC 6217.} This galaxy has a powerful bar and symmetric ring. In contrast to 
classical barred galaxies, the spiral arms in NGC~6217 begin not from the ends of the bar, 
but from the nucleus, independent of the bar. Numerous star-forming regions are observed 
in the galaxy. The nucleus is blue. The colors of the spiral arms and ring correspond to a 
stellar population with an age of 10--100~Myr, while the colors of the disk testify to active, 
ongoing star formation ($\tau_1 \approx 10$~Gyr). The bulge of NGC~6217 is shifted to 
the right of the NCS on the $(U-B)\div(B-V)$ diagram, into the region of models with an 
ongoing secondary burst of star formation. The bulge of NGC~6217 is very small (radius 
$4\arcsec-5\arcsec$), and is located between the blue nucleus and the beginning of the bar 
and spiral arms with active star formation. It is possible that we are observing in this region 
a mixture of an old stellar population in the bulge and a young population in the 
surrounding components.

{\bf NGC 6340.} This is a symmetrical lenticular galaxy. The structure, evolution, and the 
composition of the stellar population of NGC~6340 were studied in detail in \citet{chi09} 
based on photometric and spectrosopic data. \citet{chi09} detected an asymmetry in the 
composition of the stellar population of the disk of NGC~6340: the ages of stars in the 
southwest part of the disk are half the ages of those in the northeast part (5~Gyr in the 
southwest and 10~Gyr in the northeast). The nucleus, bulge, and disk occupy a region in 
the lower part of the NCS in two-color diagrams, where they are indistinguishable from 
the components of other early-type galaxies. One exception is the position of the 
NGC~6340 disk on the $(U-B)\div(B-V)$ diagram, where it is shifted upward and to the 
right of the NCS. Anomalously low values of $U-B$ are characteristic of stellar systems 
with moderate ($M_2/M_1 \sim 0.1$) secondary bursts of star formation with durations to 
several tens of millions of years. Such photometric parameters are characteristic of spiral 
arms. A system of rings (or tightly wound spiral arms) is clearly visible in the images of 
NGC~6340 presented in \citet{chi09}, obtained after subtracting the contributions of the 
bulge and disk from the image of the galaxy.

{\bf NGC 7351.} This is a symmetrical S0 galaxy with a small bar. In spite of the fact that 
NGC~7351 is the earliest-type galaxy in our sample (Table~\ref{table:tab1}), the colors 
of its components are more characteristic of Sab--Sb spiral galaxies. The nucleus of 
NGC~7351 is slightly bluer than the bulge and the disk, it has the same colors as the bar 
(Figs.~\ref{figure:fig8}a, \ref{figure:fig8}b). The disk of NGC~7351 is shifted to the right 
of the NCS on the $(U-B)\div(B-V)$ and $(B-V)\div(V-I)$ diagrams. Its color 
characteristics correspond to those of stellar systems with a secondary burst of star 
formation long in the past ($>1$~Gyr).

{\bf NGC 7678.} This is a late-type galaxy with a bar and active star formation ongoing in 
one of its spiral arms. The components are located around the NCS in the two-color 
diagrams, in the region corresponding to stellar systems with an exponentially decaying 
SFR with the characteristic time scale $\tau_1 \sim 5$~Gyr. The nucleus of NGC~7678 
deviates upward and to the right of the NCS on the $(U-B)\div(B-V)$, and to the right of 
the NCS on the $(B-V)\div(V-R)$ diagram. The color characteristics suggest a burst of 
star formation has been occurring in the nucleus over the past 10--30~Myr.

{\bf IC 1525.} This symmetric galaxy with an intermediate Sb morphological type has a 
bar and ring. Numerous star-forming complexes are observed in the spiral arms and at 
the ends of the bar \citep{bru11}. The nucleus and bar of IC~1525 are shifted to the right 
of the NCS on the $(U-B)\div(B-V)$ diagram. This provides evidence for the presence of 
an appreciable fraction of relatively young stars, together with the old stellar population.

\section{CONCLUSION}

Based on surface photometry of 26 galaxies with various morphological types (from S0 
to Sd) in the optical ($UBVRI$) and near-infrared ($JHK$), we have studied the 
compositions of the stellar populations and star-formation histories in their regular 
(nucleus, bulge, disk) and irregular (spiral arms, bar, ring) components using the 
PEGASE.2 evolutionary models. We have estimated the compositions of the stellar 
populations of the galaxy components from their positions on two-color diagrams. The 
color indices of the components have been compared with the photometric parameters 
of stellar systems derived for seven evolutionary models with an exponential decrease in 
the star-formation rate and 672 models with a secondary burst of star formation.

Our main conclusions are the following.

1. The galaxies and their components are described well by evolutionary models of stellar 
systems with zero initial metallicity, ages of 10--13~Gyr, and an exponential decrease in the 
star formation rate, with the characteristic time scale for the decay in the SFR being from 
1 to 20 billion years. The observed color indices of the galaxies and their components in 
the optical are systematically redder than the model values by $\sim0.05^m-0.1^m$.

2. A secondary burst of star formation can shift the position of a stellar system on the 
two-color diagrams to the right or left of the NCS and the absorption line. Systems with an 
ongoing burst of star formation ($T_2 < \tau_2$) are shifted to the right of the NCS on 
the $(U-B)\div(B-V)$ diagram, while systems whose secondary burst of star formation has 
ended ($T_2 > \tau_2$) are shifted to the left. In most models, a secondary burst of star 
formation shifts stellar systems to the right and/or upward relative to the NCS in the other 
two-color diagrams.

3. With uncertainties in the color indices $\approx0.1^m$, it is possible to detect a powerful 
secondary burst of star formation ($M_2/M_1 \ge 1$) occurring in the last billion years, or 
a burst with moderate power ($M_2/M_1 \approx 0.1$) occurring $\le 100-300$~Myr ago, 
from the position of a stellar system in the two-color diagrams.

4. Galaxies of early (S0--Sb) and late (Sb--Sd) types differ appreciably in the photometric 
characteristics of their components. Both galaxies of early and late types are observed along 
the Sb galaxies. S0 galaxies are not distinguished from early-type spiral galaxies in terms of 
their photometric parameters.

5. In 10 of the 26 galaxies studied, the positions of their components on the two-color 
diagrams can be explained using models with a secondary burst of star formation. The 
parameters of the inferred secondary bursts have been estimated. Among the ten galaxies 
with complex star-formation histories, five display signs of peculiarity in their morphologies.

6. In all the galaxies, the ages of the stellar populations in the spiral arms and rings did not 
exceed 1~Gyr.

\begin{acknowledgements}
This work was supported by the Russian Science Founation (project 14--22--00041). This 
work has made use of the HyperLeda (http://leda.univlyon1.fr), NED 
(http://ned.ipac.caltech.edu), and 2MASS (http://irsa.ipac.caltech.edu/applications/2MASS) 
electronic databases.
\end{acknowledgements}


\begin{thebibliography}{}

\bibitem[\protect\citeauthoryear{Arsenault}{1989}]{ars89}
Arsenault R., 1989, A\&A, 217, 66

\bibitem[\protect\citeauthoryear{Artamonov et al.}{1997}]{art97}
Artamonov B.P., Bruevich V.V., Gusev A.S., 1997, Astron. Rep., 41, 577

\bibitem[\protect\citeauthoryear{Artamonov et al.}{1999}]{art99}
Artamonov B.P., Badan Yu.Yu., Bruevich V.V., Gusev A.S., 1999, Astron. Rep., 43, 377

\bibitem[\protect\citeauthoryear{Artamonov et al.}{2000}]{art00}
Artamonov B.P., Badan Yu.Yu., Gusev A.S., 2000, Astron. Rep., 44, 561

\bibitem[\protect\citeauthoryear{Artamonov et al.}{2010}]{art10}
Artamonov B.P., Bruevich V.V., Gusev A.S. et al., 2010, Astron. Rep., 54, 767

\bibitem[\protect\citeauthoryear{Bessell \& Brett}{1988}]{bes88}
Bessell M.S., Brett J.M., 1988, PASP, 100, 1134

\bibitem[\protect\citeauthoryear{Bessell \& Murphy}{2012}]{bes12}
Bessell M., Murphy S., 2012, PASP, 124, 140

\bibitem[\protect\citeauthoryear{Bothun \& Gregg}{1990}]{bot90}
Bothun G.D., Gregg M.D., ApJ, 350, 73

\bibitem[\protect\citeauthoryear{Bruevich et al.}{2007}]{bru07}
Bruevich V.V., Gusev A.S., Ezhkova O.V., Sakhibov F.Kh., Smirnov M.A., 2007, Astron. Rep., 51, 222

\bibitem[\protect\citeauthoryear{Bruevich et al.}{2010}]{bru10}
Bruevich V.V., Gusev A.S., Guslyakova S.A., 2010, Astron. Rep., 54, 375

\bibitem[\protect\citeauthoryear{Bruevich et al.}{2011}]{bru11}
Bruevich V.V., Gusev A.S., Guslyakova S.A., 2011, Astron. Rep., 55, 310

\bibitem[\protect\citeauthoryear{Buta \& Williams}{1995}]{but95}
Buta R., Williams K.L., 1995, AJ, 109, 543

\bibitem[\protect\citeauthoryear{Carpenter}{2001}]{car01}
Carpenter J.M., 2001, AJ, 121, 2851

\bibitem[\protect\citeauthoryear{Chilingarian et al.}{2009}]{chi09}
Chilingarian I.V., Novikova A.P., Cayatte V., Combes F., di Matteo P., Zasov A.V., 2009, A\&A, 504, 389

\bibitem[\protect\citeauthoryear{de Grijs}{1998}]{gri98}
de Grijs R., 1998, MNRAS, 299, 595

\bibitem[\protect\citeauthoryear{Fioc \& Rocca-Volmerange}{1997}]{fio97}
Fioc M., Rocca-Volmerange B., 1997, A\&A, 326, 950

\bibitem[\protect\citeauthoryear{Friedli \& Benz}{1995}]{fri95}
Friedli D., Benz W., 1995, A\&A, 301, 649

\bibitem[\protect\citeauthoryear{Gavazzi et al.}{2002}]{gav02}
Gavazzi G., Bonfanti C., Sanvito G., Boselli A., Scodeggio M., 2002, ApJ, 576, 135

\bibitem[\protect\citeauthoryear{Gusev}{2006a}]{gus06a}
Gusev A.S., 2006a, Astron. Rep., 50, 167

\bibitem[\protect\citeauthoryear{Gusev}{2006b}]{gus06b}
Gusev A.S., 2006b, Astron. Rep., 50, 182

\bibitem[\protect\citeauthoryear{Gusev}{2007}]{gus07}
Gusev A.S., 2007, Astron. Rep., 51, 1

\bibitem[\protect\citeauthoryear{Gusev \& Efremov}{2013}]{gus13}
Gusev A.S., Efremov Yu.N., 2013, MNRAS, 434, 313

\bibitem[\protect\citeauthoryear{Gusev \& Kaisin}{2002}]{gus02b}
Gusev A.S., Kaisin S.S., 2002, Astron. Rep., 46, 712

\bibitem[\protect\citeauthoryear{Gusev \& Kaisin}{2004}]{gus04}
Gusev A.S., Kaisin S.S., 2004, Astron. Rep., 48, 611

\bibitem[\protect\citeauthoryear{Gusev \& Park}{2003}]{gus03b}
Gusev A.S., Park M.-G., 2003, A\&A, 410, 117

\bibitem[\protect\citeauthoryear{Gusev et al.}{2002}]{gus02a}
Gusev A.S., Zasov A.V., Kaisin S.S., Bizyaev D.V., 2002, Astron. Rep., 46, 704

\bibitem[\protect\citeauthoryear{Gusev et al.}{2003}]{gus03a}
Gusev A.S., Zasov A.V., Kaisin S.S., 2003, Astron. Lett., 29, 363

\bibitem[\protect\citeauthoryear{Hodge}{1982}]{hod82}
Hodge P.W., 1982, AJ, 87, 1341

\bibitem[\protect\citeauthoryear{Jarrett et al.}{2000}]{jar00}
Jarrett T.H., Chester T., Cutri R., Schneider S., Skrutskie M., Huchra J.P., 2000, AJ, 119, 2498

\bibitem[\protect\citeauthoryear{Kinman \& Castelli}{2002}]{kin02}
Kinman T., Castelli F., 2002, A\&A, 391, 950

\bibitem[\protect\citeauthoryear{Landolt}{1992}]{lan92}
Landolt A.U., 1992, AJ, 104, 340

\bibitem[\protect\citeauthoryear{Martinet \& Friedli}{1997}]{mar97}
Martinet L., Friedli D., 1997, A\&A, 323, 363

\bibitem[\protect\citeauthoryear{M\"{o}llenhoff}{2004}]{mol04}
M\"{o}llenhoff C., 2004, A\&A, 415, 63

\bibitem[\protect\citeauthoryear{Mosenkov et al.}{2010}]{mos10}
Mosenkov A.V., Sotnikova N.Ya., Reshetnikov V.P., 2010, MNRAS, 401, 559

\bibitem[\protect\citeauthoryear{Salpeter}{1955}]{sal55}
Salpeter E.E., 1955, ApJ, 121, 161

\bibitem[\protect\citeauthoryear{Sandage}{1986}]{san86}
Sandage A., 1986, A\&A, 161, 89

\bibitem[\protect\citeauthoryear{Schlafly \& Finkbeiner}{2011}]{sch11}
Schlafly E.F., Finkbeiner D.P., 2011, ApJ, 737, 103

\bibitem[\protect\citeauthoryear{Schr\"{o}eder \& Visvanathan}{1996}]{sch96}
Schr\"{o}eder A., Visvanathan N., 1996, A\&AS, 118, 441

\bibitem[\protect\citeauthoryear{Young et al.}{1996}]{you96}
Young J.S., Allen L., Kenny J.D.P., Lesser A., Rownd B., 1996, AJ, 112, 1903

\bibitem[\protect\citeauthoryear{Zasov \& Sil'chenko}{1983}]{zas83}
Zasov A.V., Sil'chenko O.K., 1983, Sov. Astron., 27, 616

\end{thebibliography}
\end{document}